\documentclass[prl,aps,groupedaddress,twocolumn,showpacs,floatfix,preprintnumbers,superscriptaddress]{revtex4-1}
\usepackage{amsmath,amssymb,multirow,bm}

\newcommand{\fig}[1]{Fig.~\ref{#1}}

\newcommand{\beq}{\begin{equation}}
\newcommand{\eeq}{\end{equation}}
\newcommand{\bea}{\begin{eqnarray}}
\newcommand{\eea}{\end{eqnarray}}

\newcommand{\benn}{\begin{displaymath}}
\newcommand{\eenn}{\end{displaymath}}

\usepackage{graphicx}
\usepackage[breaklinks=true]{hyperref}
\usepackage[update,prepend]{epstopdf}
\usepackage{color}

\begin{document}
\preprint{LA-UR-14-21210, NT@UW-14-06}

\title{Relativistic Coulomb excitation within Time Dependent Superfluid Local Density Approximation}

\author{I. Stetcu}
\affiliation{Theoretical Division, Los Alamos National Laboratory, Los Alamos, NM 87545, USA}
\author{C. A. Bertulani}
\affiliation{Department of Physics and Astronomy, Texas A \& M University - Commerce, Commerce, TX 75429, USA}
\author{A. Bulgac}
\affiliation{Department of Physics, University of Washington, Seattle, WA 98195--1560, USA}
\author{P. Magierski}
\affiliation{Department of Physics, University of Washington, Seattle, WA 98195--1560, USA}
\affiliation{Faculty of Physics, Warsaw University of Technology,
ulica Koszykowa 75, 00-662 Warsaw, POLAND}
\author{K.J. Roche}
\affiliation{Department of Physics, University of Washington, Seattle, WA 98195--1560, USA}
\affiliation{Pacific Northwest National Laboratory, Richland, WA 99352, USA}

\begin{abstract}
Within the framework of the unrestricted time-dependent
density functional theory, we present for the first time an analysis
of the relativistic Coulomb excitation of the heavy deformed open shell 
nucleus $^{238}$U.  The approach is based on Superfluid Local Density
Approximation (SLDA) formulated on a spatial lattice that can take
into account coupling to the continuum, enabling self-consistent
studies of superfluid dynamics of any nuclear shape.  We have computed
the energy deposited in the target nucleus as a function of the impact
parameter, finding it to be significantly larger than the estimate
using the Goldhaber-Teller model. The isovector giant dipole resonance, 
the dipole pygmy resonance and giant quadrupole modes were excited 
during the process.  The one body dissipation of collective dipole 
modes  is shown to lead a damping width $\Gamma_\downarrow \approx 0.4$ 
MeV and the number of pre-equilibrium neutrons emitted has been quantified.

\end{abstract}

\date{\today}


\maketitle


Coulomb excitation represents an ideal method to probe the properties
of large amplitude nuclear motion, because the excitation process is
not obscured by uncertainties related to nuclear forces.  The
excitation probabilities are governed by the strength of the Coulomb
field only and they can be fully expressed in terms of the
electromagnetic multipole matrix elements \cite{wa, bb,
emling,*abe,bp,lanza,hussein}.  From the theoretical point of view,
Coulomb excitation can be treated as a textbook example of a nuclear
system being subjected to an external, time-dependent perturbation.
However, in order to be able to probe nuclear collective modes
involving multi-phonon states for example \cite{Boretzky,Ilievski}, a large amount of energy
has to be transferred to the nuclear system. Thus the interaction time
should be relatively short and the velocity of the projectile has to
be sufficiently large for an efficient excitation of nuclear modes of
frequency $\omega$, the collision time $\tau_{coll}=b/\gamma v$ has to
fulfill the condition that $\omega\tau_{coll} \simeq 1$.  Here $b$ is
the impact parameter, $v$ is the projectile velocity, and $\gamma =
(1-v^2/c^2)^{-1/2}$ is the Lorentz factor.  One of the best known
examples of collective nuclear motion is the isovector giant dipole
resonance (IVGDR).  A reasonably good estimate of the IVGDR
vibrational frequency is $\hbar \omega \approx 80 {\rm MeV}/A^{1/3}$ for
spherical nuclei.  It implies that the excitation of a heavy nucleus
to such energies requires a relativistic projectile.  

We report on a new and powerful method to study relativistic
Coulomb excitation and nuclear large amplitude collective 
motion in the framework of Time Dependent Superfluid Local Density 
Approximation (TDSLDA). This is a fully microscopic approach to the
problem based on an extension of the Density Functional Theory (DFT)
to superfluid nuclei and time-dependent external probes, where all the
nuclear degrees of freedom are taken into account on the same footing,
without any restrictions and where all symmetries (translation,
rotation, parity, local Galilean covariance, local gauge symmetry, isospin symmetry,
minimal gauge coupling to electromagnetic (EM) fields) are correctly implemented \cite{bulgac,
stetcu}. The interaction between the impinging $^{238}$U projectile
and the $^{238}$U target is very strong ($\propto Z_p Z_t \alpha
\approx 62$, where $\alpha$ is the fine structure constant), which
thus require a non-perturbative treatment, and the excitation process
is highly non-adiabatic.  We assume a completely classical
projectile straight-line motion since its de Broglie wavelength is of
the order of $0.01$ fm for $\gamma \sim 1.5-2$. In evaluating the
EM-field created by the uranium projectile with a constant velocity
$v=0.7c$ along the z-axis, we neglect its deformation. The projectile
produces an EM-field described by scalar 
and vector
Lienard-Wiechert potentials. 
These fields couple to a
deformed $^{238}$U target nucleus residing on a spatial lattice, see
Ref. \cite{suppl}. The interaction leads to a CM motion of the target as
well as to its internal excitation and full 3D dynamical deformation 
of the target. In order to follow the internal
motion for a long enough trajectory that allows the extraction of
useful information, we perform a transformation to a system in which
the lattice moves with the CM. The required transformation for each
single particle wave function reads $ \phi_n(\mathbf{r}, t) =
\exp(i\mathbf{R}(t)\cdot\mathbf{ \hat{p}})\psi_n(\mathbf{r}, t)$, with
$\bf{R}(t)$ describing the CM motion and $ {\bf \hat{p}}$ the momentum
operator.  The equation of motion (simplified form here) for $\phi_n$
becomes
 \bea
i\hbar  \dot{\phi}_n(\mathbf{r}, t)  = \left [ \hat{H}\left({\bf r}+{\mathbf R}(t),t\right) + 
                  \dot{{\bf R}}(t) \cdot {\hat{\bf p}} \right ]\phi_n(\mathbf{r}, t),
                   \label{eq:schr}
\eea
where $\dot{{\bf R}}(t)=\int d^3r\: \bm{j}(\bm{r},t)/M$ is the CM
velocity and $\bm{j}(\bm{r},t)$ the total current density.

The target nucleus is described within the SLDA and its time evolution
is governed by the TD meanfield-like equations (spin degrees of freedom are not shown): 
\bea
\lefteqn{i\hbar\frac{\partial}{\partial t} 
\left  ( \begin{array} {c}
  U({\bf r},t)\\  
  V({\bf r},t)\\ 
\end{array} \right ) \nonumber }\\
&& =
\left ( \begin{array}{cc}
h({\bf r},t)&\Delta({\bf r},t)\\
\Delta^*({\bf r},t)&-h^* ({\bf r},t)
\end{array} \right )  
\left  ( \begin{array} {c}
  U({\bf r},t)\\
  V({\bf r},t)\\ 
\end{array} \right ).
\eea
The single-particle Hamiltonian $h(\mathbf{r},t)$ and the pairing
field $\Delta({\bf r},t)$ are obtained self-consistently from an
energy functional that is in general a function of various normal,
anomalous, and current densities. The external electromagnetic (EM) field
has the minimal gauge coupling $\mbox{{\boldmath{$\nabla$}}}_{A} =
\mbox{{\boldmath{$\nabla$}}} - i{\bf A}/{\hbar c} $ (and similarly
for the time-component) in all terms with currents, as well as in the
definition of the momentum operator $ \mathbf{\hat p}$ in
Eq. (\ref{eq:schr}), details in \cite{suppl}.  In the current
calculation, the Skyrme SLy4 energy functional \cite{SLy4,*Sly4E} was
adopted, with nuclear pairing as introduced in Ref. \cite{yu2003}, 
which provides a very decent description of the IVGDR in
$^{238}$U \cite{stetcu}. The coupling between the spin and the
magnetic field was neglected. The Coulomb self-interaction between
protons of the target nucleus is taken into account using the
modification of the method described in Ref. \cite{castro2003}, so as not to 
include contributions from images in neighboring cells. For
the description of the numerical methods see Refs.
\cite{kenny,Bulgac2011} and many other technical details can be found in
\cite{suppl}.

The DFT approach to quantum dynamics has some peculiar
characteristics.
Unlike a
regular quantum mechanics (QM) treatment one does not have access to
wave functions, but instead to various one-body densities and
currents. Within a DFT approach quantities
trivial to evaluate in QM 
become basically impossible to calculate. For example,
by solving the Schr\"odinger equation one can evaluate at any time the
probability that a system remained in its initial state from
${\cal P}(t)=|\langle \Psi(0)|\Psi(t)\rangle |^2$, where $\Psi(t)$ is the
solution of the Schr\"odinger equation (or some of its
approximations).  Within 
DFT
one has access to the one-body (spin-)density $
\rho(\bm{r},t)$ and one-body (spin-)current $\bm{j}(\bm{r},t)$ 
with no means to compute the probability ${\cal P}(t)$.
One can calculate for example a quantity such as $\int
d^3r \rho(\bm{r},0)\rho(\bm{r},t)$, but there is no obvious way to
relate it to the probability ${\cal P}(t)$. One might try to define ${\cal P}(t)$
instead through the overlap of the initial and current ``Slater
determinants" constructed through the fictitious single-particle wave
functions entering the DFT formalism, which is a rather arbitrary
postulate. One can find quite often in literature various formulas
used within DFT treatment of nuclei, which are simply ``copied" from
various mean field approaches, without any solid justification
provided. 
Restrictions inherent to a DFT
approach, prevent us from being able to calculate various quantities,
which within a QM approach are easy to evaluate. Even though we
evaluate accurate densities and currents well beyond the linear regime, within a
DFT approach we cannot separate for example the emission of one and
two photons from an excited nucleus, which however could be estimated
relative easily within a perturbative linear response approach such a
(Q)RPA. On the other hand a DFT approach has unquestionable
advantages, allowing us to go far into the non-linear regime and
describe large amplitude collective motion.

\begin{figure} 
\includegraphics*[width=\columnwidth]{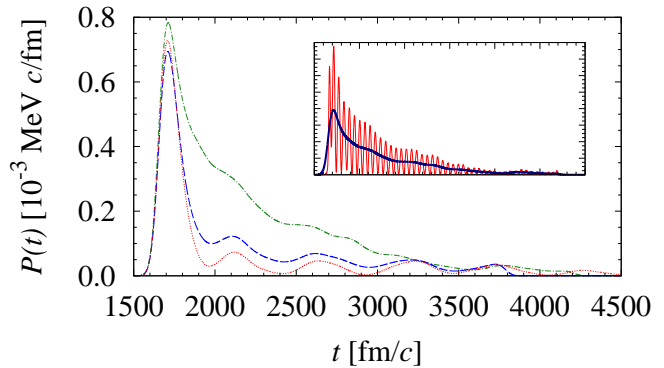}
\caption{ (color online) 
The emitted energy rate via EM radiation for a collision with impact parameter $b=16.2$ fm, for three orientations. In two cases the  nuclear symmetry axis is parallel to the reaction plane and perpendicular (dot-dashed line)  or parallel (dashed line) with respect to the incoming projectile, while in the third it is both perpendicular to the reaction plane and the incoming projectile (dotted line). These configurations are denoted by $\perp\parallel$, $\parallel$ and $\perp\perp$, respectively. We show time-averaged quantities, while in the insert, for one configuration, we also show the raw, strongly oscillating, data. The rate at which this quantity changes is directly related to the characteristic damping time, which we estimate at $500 \ \mathrm{fm}/c$, leading to a width $\Gamma_\downarrow\approx 0.4$ MeV.
}\label{radiation}
\end{figure}

The incoming projectile excites various modes in the target
nucleus and the axial symmetry of the initial ground state is lost. 
Because $^{238}$U is highly deformed the energy of the first $2^+$
is 45 keV, corresponding to a very long rotational period, and thus 
during simulation time  considered here ($\approx 10^{-20}$ sec.)
it can be considered fixed.
The identification of these modes requires certain care, since
during the collision the system is beyond the linear regime and the
analysis using the response function is not applicable in general.
However, the information about the excited nuclear modes is carried in
the subsequent EM radiation leading to nuclear
de-excitation.  De-excitation to the ground state via photon emission
requires times of about $10^{-16}$ sec., which is four orders
of magnitude longer than in the current calculations. 
However,  it is possible to compute the spectrum
of the pre-equilibrium neutrons and gamma radiation, which allows the
identification of the excited nuclear modes. We can accurately treat the
system as a classical source of electromagnetic radiation and the time
dependence of the proton current governs the rate of emission, see
Refs. \cite{suppl,Baran1996,Oberacker2012}:
\bea
P(t + r/c) &=& \frac{e^{2}}{\pi c } \sum_{l,m} \left | \int_{-\infty}^{\infty} {\bf b}_{lm}(k,\omega) e^{-i \omega t} d\omega \right |^{2} , \label{power}
\eea 
with
${\bf b}_{lm}(k,\omega) = \int dt \,d^{3}r \,e^{-i \omega t} \mbox{{\boldmath{$\nabla$}}}\times {\bf j}({\bf r},t) j_{l}(kr)Y_{lm}^{*}(\hat{\bf r})$. 
Here $\omega=kc$, $j_l(kr)$ is the spherical Bessel function of order $l$, and ${\bf j}({\bf r},t)$ is the proton current.  The
emission rate $P$ is plotted in Fig. \ref{radiation}.  The magnitude of
this quantity indicates that the total amount of radiated energy
during the evolution time (about $2500$ fm/c) is rather small compared
to the total absorbed energy and does not exceed $1$ MeV, which is
about $2-3\%$ of the deposited energy reported in Table
\ref{table:results} below.  This implies that the effect of damping of
nuclear motion due to the emitted radiation can be neglected for
such short time intervals. Consequently, the decreasing intensity of
the radiation, see Fig. \ref{radiation},  is merely related to the rearrangements of the intrinsic
structure of the excited nucleus caused by damping of collective modes
due to the one-body dissipation mechanism. It has to be emphasized that within the
framework of the presented approach one is able to extract only a small
fraction of the spreading width $\Gamma_\downarrow$, which is due to the one-body 
dissipation mechanism. The two-body effects require e.g. stochastic extension of TDSLDA
which would allow for a dynamic hopping between various mean-fields, 
and thus could account for collisional damping as well.
 
\begin{figure}[htb] 
\includegraphics*[width=\columnwidth]{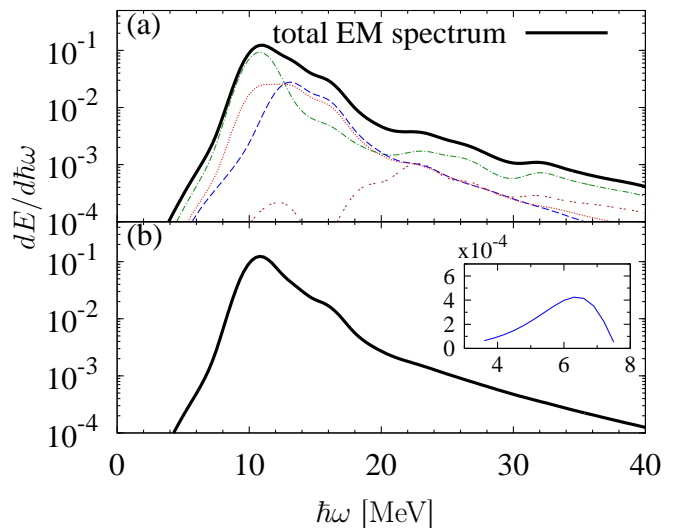}
\caption{ (color online) (a) The total energy spectrum (solid line) of emitted
EM radiation, averaged over the target-projectile configurations, at the impact parameter $b=12.2$ fm.  We show the total quadrupole contribution (double-dotted line), as well as the contributions 
from the three target-projectile orientations using the same symbols as in Fig. \ref{radiation}. 
(b) The radiation emitted from the target nucleus when only the dipole component of the projectile
electromagnetic field is used. 
The insert shows the pygmy resonance contribution to the emitted
spectrum, visible in the main figure as the slope change
at low energies. }\label{photons}
\end{figure}

TDSLDA provides the EM power spectrum 
\cite{suppl,Baran1996,Oberacker2012}, 
$
\frac{dE}{d\omega}
={{4 e^2}\over{c}} \sum_{l,m} |{\bf b}_{lm}(k,\omega)|^{2} , \label{spectrum}
$ 
arising from the multipole expansion in Eq.  (\ref{power}).  
This quantity is different from what one would
compute within a linear response approach or first order perturbation theory, 
see, e.g., Refs. \cite{wa,bb,emling,*abe,bp,lanza,hussein}, which provides the excitation 
probability only $\propto |\int d{^3}r \rho_{tr}(\bm{r})V_{ext}(\bm{r})|^2$, where $\rho_{tr}(\bm{r})$ 
is the transition density and $V_{ext}(\bm{r})$ - the external field. $dE/d\omega$ is 
proportional to the excitation probability, here in the non-linear regime, and the subsequent photon
emission probability as well. A typical example of the
emitted EM radiation for a given impact parameter is shown in
Fig. \ref{photons}, here due to the internal excitation of the
system alone. The EM radiation due the CM motion has been calculated separately (see Table
\ref{table:results} and \cite{suppl}).

In Fig. \ref{photons}(a) the emitted radiation shows a well defined
maximum at energy $10-12$ MeV which corresponds to the excitation of
IVGDR. We have applied a smoothing of the original calculations using
the half-width of $1$ MeV. Therefore, the original separate peaks split
due to the deformation of $^{238}$U merge into a single wider peak.
However, at larger frequencies another local maximum exists which we
associate with the isovector giant quadrupole resonance (IVGQR). In
order to rule out other possibilities we have repeated the calculation
by retaining only the dipole component of the electromagnetic field
produced by the projectile \cite{suppl}.  The results are shown in  
Fig. \ref{photons}(b). In this case, the high-energy
structure above 20 MeV disappears, evidence that
the high energy peak is related to the IVGQR. Noticeable contribution to the total radiation is
coming from the quadrupole component of radiated field.

At low energies a change of slope
occurs at about $\hbar\omega \approx 7$ MeV, present at
the same energy for all impact parameters and orientations, see
Ref. \cite{suppl}.  It indicates 
a considerable amount of strength at low energies, giving rise to an
additional contribution to the EM radiation. We attribute
this additional structure to the excitation of the pygmy dipole
resonance (PDR). The inset of the figure \ref{photons} shows the spectrum of
emitted radiation due to this mode. The contribution to the total
radiated energy coming from the PDR is rather small and
reads: $1.7$, $2.4$, $1.5$ and $0.8$ keV for impact parameters $12.2$, $14.6$, $16.2$
and $20.2$ fm, respectively. It corresponds to about $0.22$\%,
$0.5$\%, $0.43$\%, $0.45$\% of the emitted radiation (due to internal motion)
respectively. The relatively 
small amount of E1 strength obtained in our calculations, in the region where 
the PDR is expected, agrees with recent measurements \cite{pygmy}. 

The comparison between the average energy transferred to the internal motion
of the target nucleus for three values of the impact parameter
obtained within TDSLDA and within a simplified Goldhaber-Teller
model \cite{Goldhaber-Teller} presented in Table \ref{table:results} shows
that significantly more energy is deposited by the projectile within
the TDSLDA. The Goldhaber-Teller model is equivalent to a linear response 
result, assuming that all isovector transition strength is concentrated in
two sharp lines, corresponding to an axially deformed target. 
An exact QRPA approach would therefore severely
underestimate the amount of internal energy deposited, one reason being 
the non-linearity of the response, naturally incorporated in TDSLDA.  
Another reason is the fact that the present microscopic framework
describing the target allows for many degrees of freedom to be excited, apart
from pure dipole oscillations.  
At the same time, the CM target energy  
alone is approximately the same as obtained in a simplified point particles
Coulomb recoil model of both the target and projectile.

\begin{table}
\caption{
Internal excitation energy in TD-SLDA ($E_{int}$)  and in the Goldhaber-Teller model ($E_{GT}$), as well as the EM energy radiated ($E^{int}_\gamma$) from the excited nucleus during time interval $\delta t = 2500$ fm/c after collision, for four values of impact parameters $b$ and three orientations of the nucleus with respect to the beam. We also list their respective ratios to the total transferred energy. Finally, the Goldhaber-Teller model results ($E^{*}_{GT}$) for $m^{*}=0.7m$ effective mass are presented in the last column. All energies are in MeV. 
}

\label{table:results}
\begin{ruledtabular}
\begin{tabular}{ccccccc} 
$b (fm)$   & $E_{int}$ &$E_{int}/E$  & $E^{int}_\gamma$ &  $E^{int}_\gamma/E_{\gamma}$ & $E_{GT}$ & $E^{*}_{GT}$ \\
\hline
12.2 $\perp ||$   & 39.29 & 0.668 &  0.911  &  0.960 & 17.58 & 24.68 \\
14.6 $\perp ||$   & 19.2  & 0.608 &  0.567  &  0.963 & 10.32 & 14.51 \\
16.2 $\perp ||$   & 12.87 & 0.547 &  0.411  &  0.963 &  7.41 & 10.43 \\
20.2 $\perp ||$   & 5.41  & 0.444 &  0.199  &  0.961 &  3.43 &  4.84 \\

12.2 $||$         & 25.11 & 0.588 &  0.5    &  0.941 & 12.94 & 18.17 \\
14.6 $||$         & 13.16 & 0.498 &  0.306  &  0.942 &  7.22 & 10.16 \\
16.2 $||$         & 8.97  & 0.470 &  0.217  &  0.939 &  5.02 &  7.07 \\
20.2 $||$         & 3.8   & 0.367 &  0.106  &  0.934 &  2.16 &  3.05 \\

12.2 $\perp\perp$ & 24.21 & 0.591 &  0.407  &  0.930 & 12.36 & 17.33 \\
14.6 $\perp\perp$ & 12.58 & 0.513 &  0.245  &  0.929 &  6.65 &  9.34 \\
16.2 $\perp\perp$ &  8.5  & 0.464 &  0.175  &  0.926 &  4.49 &  6.32 \\
20.2 $\perp\perp$ &  3.5  & 0.353 &  0.085  &  0.919 &  1.78 &  2.51 \\
\end{tabular}
\end{ruledtabular}

\end{table} 

The average energy radiated due to the internal excitation represents only
part of the total radiated energy. (One should remember that a straightforward 
DFT approach provides no measure for the variance.) Also, because of the 
spreading of the strength due to one-body dissipation only a fraction of the energy
$\Gamma_\gamma/\Gamma$ (where $\Gamma_\gamma$ is the EM-width alone 
and $\Gamma$ the total width of the IVGDR) is emitted as a pulse, as shown 
in \fig{radiation}. A subsequent pulse of reduced amplitude is to be expected 
after a delay $\approx  \Gamma/(\Gamma_\gamma \omega_{IVGDR}) \approx 10^5 \ldots 10^6$ fm/c. 
Since our simulations times are much shorter we are not able to 
see emission of the second photon, as reported in experiment \cite{Boretzky,Ilievski},
where two photons were measured in coincidence.
In our calculations we have followed the nuclear evolution
during approximately $2500$ fm/c after collision. 
The other component of the EM radiation arises from the
CM acceleration as a result of collision (Bremsstrahlung),
during the relatively short time interval $\tau_{coll}={b}/{v\gamma}$.
The radiation emitted from the internal motion has a much longer time scale. 

We can estimate the cross section for the emission of radiation by
assuming that the asymptotic transition probability for a given impact parameter $b$ is given by ${\cal P}\propto \frac{1}{3} (\frac{E_{\gamma\perp\perp}(b)}{E_{\perp\perp}(b)} + 
       \frac{E_{\gamma\perp ||}(b)}{E_{\perp ||}(b)} + 
       \frac{E_{\gamma ||}(b)}{E_{||}(b)})$.
Here $E_{\perp\perp}(b)$, $E_{\perp ||}(b)$ and $E_{||}(b)$ are the total energies
transferred to the target nucleus during the collision at the impact
parameter $b$ and for the three independent orientations. 
Our simulation yields $\sigma_{\gamma}=2\pi\int {\cal P} bdb \simeq 108$ mb.
A detailed comparison of intensities of radiation for
various impact parameters and orientations is shown in Table
\ref{table:results}. It is evident that although the intensity of
radiation decreases with increasing impact parameter, the ratio between
the intensities due to the internal modes with that of the CM motion
remains fairly constant and depends slightly on orientation
For the orientation perpendicular to the beam
and parallel to the reaction plane the target
nucleus is the most efficiently excited which results in a larger ratio.

\begin{figure}
\includegraphics*[width=\columnwidth]{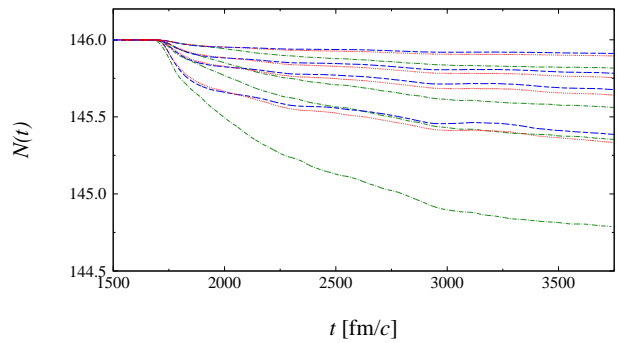}
\caption{ (color online) The number of neutrons inside the sphere of
radius $15$ fm around the target nucleus as a function of time for the four impact parameters. 
We use the same convention as in Fig. \ref{radiation} for the possible orientations.
The emission rate is inverse proportional with the value of the impact parameter.}
\label{neutrons}
\end{figure}

The average energy deposited in the target nucleus is of the order of
the neutron separation energy.  In Fig. \ref{neutrons}
we plot the total number of neutrons inside a (smoothed) sphere of
radius $15$ fm which is slightly larger than the nuclear diameter (see
Ref. \cite{suppl} for details).  For all these impact parameters neutrons
can leak from the excited system. Since more energy is deposited in
the nucleus with perpendicular orientation with respect to the beam,
the rate of emitted neutrons is larger in that case. However, the simulation box is large enough (about 40 times bigger than the nucleus) so that during the evolution the calculations are not affected by the emitted neutrons.\\


We thank G.F. Bertsch for a number of discussions and reading the manuscript. 
We acknowledge support under U.S. DOE Grants DE-FC02-07ER41457, DE-
FG02-08ER41533,  NSF grant PHY-1415656, Polish National Science Centre (NCN) Grant, decision 
no. DEC-2013/08/A/ST3/00708, and in part by the ERANET-NuPNET grant SARFEN
    of the Polish National Centre for Research and Development (NCBiR). Part of this work was performed under the 
auspices of the National Nuclear Security Administration of the US Department 
of Energy at Los Alamos National Laboratory under contract No. DE-AC52-06NA25396. 
This research used resources of the National Energy Research Scientific Computing 
Center, which is supported by the Office of Science of the U.S. Department of Energy 
under Contract No. DE-AC02-05CH11231 and of the Oak Ridge Leadership Computing 
Facility at the Oak Ridge National Laboratory, which is supported by the Office of Science 
of the U.S. Department of Energy under Contract No. DE-AC05-00OR22725. Some of 
the calculations reported here have been performed at the University of Washington 
Hyak cluster funded by the NSF MRI Grant No. PHY-0922770. KJR was supported 
by the DOE Office of Science, Advanced Scientific Computing Research, under 
award number 58202 "Software Effectiveness Metrics" (Lucille T. Nowell). 

\bibliography{../../bibliography}

\newpage

\onecolumngrid

\vspace{0.75cm}

\noindent{\large \bf Supplemental material to:\\ \\
Relativistic Coulomb excitation within Time Dependent Superfluid Local Density Approximation}

\vspace{0.75cm}

\noindent{\large I. Stetcu, C. Bertulani, A. Bulgac, P. Magierski and K.J. Roche}

\date{\today}

\vspace{1cm}

\begin{center}
{\bf Density functional in TDSLDA and coupling to electromagnetic field}
\end{center}

Here we present various definitions and conventions which we have
used in the manuscript. The density functional is constructed from the following local densities
and currents:

\begin{itemize}
\item density: $\rho({\bf r})=\rho({\bf r},{\bf r'})|_{r=r'}$
\item spin density: $\vec{s}({\bf r})=\vec{s}({\bf r},{\bf r'})|_{r=r'}$
\item current: $\vec{j}({\bf r})=\frac{1}{2i}(\vec{\nabla}-\vec{\nabla}')\rho({\bf r},{\bf r'})|_{r=r'}$
\item spin current (2nd rank tensor): ${\bf J}({\bf r})=\frac{1}{2i}(\vec{\nabla}-\vec{\nabla}')\otimes\vec{s}({\bf r},{\bf r'})|_{r=r'}$
\item kinetic energy density: $\tau({\bf r})=\vec{\nabla}\cdot\vec{\nabla}'\rho({\bf r},{\bf r'})|_{r=r'}$
\item spin kinetic energy density: $\vec{T}({\bf r})=\vec{\nabla}\cdot\vec{\nabla}'\vec{s}({\bf r},{\bf r'})|_{r=r'}$
\item anomalous density: $\chi({\bf r}) = \chi({\bf r}\sigma, {\bf r'}\sigma')|_{r=r',\sigma=1,\sigma'=-\sigma}$
\end{itemize}
where 
\begin{eqnarray}
\rho({\bf r},{\bf r'})&=&\sum_{\mu}
\left ( V_{\mu}^{*}({\bf r} +)V_{\mu}({\bf r'} +) +
        V_{\mu}^{*}({\bf r} -)V_{\mu}({\bf r'} -)
 \right ) \nonumber \\
s_{x}({\bf r},{\bf r'})&=&\sum_{\mu}
\left ( V_{\mu}^{*}({\bf r} +)V_{\mu}({\bf r'} -) +
        V_{\mu}^{*}({\bf r} -)V_{\mu}({\bf r'} +)
 \right ) \nonumber  \\
s_{y}({\bf r},{\bf r'})&=&i\sum_{\mu}
\left ( V_{\mu}^{*}({\bf r} +)V_{\mu}({\bf r'} -) -
        V_{\mu}^{*}({\bf r} -)V_{\mu}({\bf r'} +)
 \right )  \nonumber \\
s_{z}({\bf r},{\bf r'})&=&\sum_{\mu}
\left ( V_{\mu}^{*}({\bf r} +)V_{\mu}({\bf r'} +) -
        V_{\mu}^{*}({\bf r} -)V_{\mu}({\bf r'} -)
 \right )  \nonumber \\
\tau({\bf r},{\bf r'})&=&\sum_{\mu}
\left ( \vec{\nabla}V_{\mu}^{*}({\bf r} +)\cdot\vec{\nabla}V_{\mu}({\bf r'} +) +
        \vec{\nabla}V_{\mu}^{*}({\bf r} -)\cdot\vec{\nabla}V_{\mu}({\bf r'} -)
 \right )  \nonumber \\
\vec{j}({\bf r})&=&- Im\left ( \sum_{\mu}   
            V_{\mu}^{*}({\bf r} +)\cdot\vec{\nabla}V_{\mu}({\bf r} +) +
            V_{\mu}^{*}({\bf r} -)\cdot\vec{\nabla}V_{\mu}({\bf r} -)\right )  =  \nonumber \\
            &=& Im\left ( \sum_{\mu}   
            V_{\mu}({\bf r} +)\cdot\vec{\nabla}V_{\mu}^{*}({\bf r} +) +
            V_{\mu}({\bf r} -)\cdot\vec{\nabla}V_{\mu}^{*}({\bf r} -)
  \right )  \nonumber \\
J_{x}({\bf r})& =& Im\left ( 
  \frac{\partial}{\partial y}s_{z}({\bf r},{\bf r'})
 -\frac{\partial}{\partial z}s_{y}({\bf r},{\bf r'})
\right )|_{r=r'}  \nonumber \\
J_{y}({\bf r})& = &Im\left ( 
  \frac{\partial}{\partial z}s_{x}({\bf r},{\bf r'})
 -\frac{\partial}{\partial x}s_{z}({\bf r},{\bf r'})
\right )|_{r=r'}  \nonumber \\
J_{z}({\bf r})& =& Im\left ( 
  \frac{\partial}{\partial x}s_{y}({\bf r},{\bf r'})
 -\frac{\partial}{\partial y}s_{x}({\bf r},{\bf r'})
\right )|_{r=r'}
\chi({\bf r}\sigma, {\bf r'}\sigma')=
      \sum_{\mu}V_{\mu}^{*}({\bf r}\sigma)U_{\mu}({\bf r'}\sigma') 
\end{eqnarray}

The coupling of the nuclear system to the electromagnetic field:
\begin{eqnarray}
\vec{E} &=& -\vec{\nabla}\phi - \frac{1}{c}\frac{\partial\vec{A}}{\partial t} \\
\vec{B} &=& \vec{\nabla}\times\vec{A} \\
\end{eqnarray}
requires the following transformation of proton densities and currents:
\begin{itemize}
\item density: $\rho_{A}({\bf r})=\rho_{A}({\bf r})$
\item spin density: $\vec{s}_{A}({\bf r})=\vec{s}({\bf r})$
\item current: $\vec{j}_{A}({\bf r})=\vec{j}({\bf r}) - \frac{e}{\hbar c}\vec{A}\rho({\bf r})$
\item spin current (2nd rank tensor): ${\bf J}_{A}({\bf r})={\bf J}({\bf r}) - \frac{e}{\hbar c}\vec{A}\otimes\vec{s}({\bf r})$
\item spin current (vector): $\vec{J}_A({\bf r})  = \vec{J}({\bf r}) -\frac{e}{\hbar c} \vec{A} \times \vec{s}({\bf r})$ 
\item kinetic energy density: $\tau_{A}({\bf r})=
\left  (\vec{\nabla} - i\frac{e}{\hbar c}\vec{A}\right)\cdot\left (\vec{\nabla}' + i\frac{e}{\hbar c}\vec{A}\right)\rho({\bf r},{\bf r'})|_{r=r'} \\
 = \tau({\bf r}) -2\frac{e}{\hbar c}\vec{A}\cdot\vec{j}({\bf r}) +\frac{e^2}{\hbar^2 c^2}|\vec{A}|^2\rho({\bf r})   
 = \tau({\bf r}) -2\frac{e}{\hbar c}\vec{A}\cdot\vec{j}_A({\bf r}) -\frac{e^2}{\hbar^2 c^2}|\vec{A}|^2\rho({\bf r})$
\item spin kinetic energy density: $\vec{T}_A({\bf r})=
\left (\vec{\nabla} - i\frac{e}{\hbar c}\vec{A}\right)\cdot\left (\vec{\nabla}' + i\frac{e}{\hbar c}\vec{A}\right)\vec{s}({\bf r},{\bf r'})|_{r=r'}\\
 =\vec{T}({\bf r}) -2\frac{e}{\hbar c}\vec{A}^T \cdot {\bf J}({\bf r}) +\frac{e^2}{\hbar^2 c^2}|\vec{A}|^2\vec{s}({\bf r}) 
 =\vec{T}({\bf r}) -2\frac{e}{\hbar c}\vec{A}^T \cdot {\bf J}_A({\bf r}) -\frac{e^2}{\hbar^2 c^2}|\vec{A}|^2\vec{s}({\bf r})$
\end{itemize}

As a result the proton single particle hamiltonian has the form:
\begin{eqnarray}
h_{A}({\bf r})= -\vec{\nabla}_{A}\cdot \left  ( B({\bf r}) + \vec{\sigma}\cdot\vec{C}({\bf r}) \right )\vec{\nabla}_{A} + U_{A}({\bf r})
+ \frac{1}{2i}\left ( \vec{W}({\bf r})\cdot( \vec{\nabla}_A\times\vec{\sigma} ) + 
\vec{\nabla}_A\cdot( \vec{\sigma}\times\vec{W}({\bf r}) ) \right ) \nonumber \\
 +\vec{U}_{\sigma}^{A}({\bf r})\cdot\vec{\sigma} 
+\frac{1}{i} \left ( \vec{\nabla}_{A}\cdot\vec{U}_{\Delta}^{A}({\bf r}) + \vec{U}_{\Delta}^{A}({\bf r})\cdot\vec{\nabla}_{A} \right )
\end{eqnarray}
and
\begin{eqnarray}
&&U_{A}({\bf r}) =  U({\bf r}) - C^{\nabla J} \frac{e}{\hbar c}  \vec{\nabla}\cdot[\vec{A}\times\vec{s}({\bf r})] 
-C^{\tau}\left ( 2\frac{e}{\hbar c} \vec{A}\cdot \vec{j}({\bf r})+\frac{e^2}{\hbar^2 c^2}|\vec{A}|^2\rho({\bf r}) \right ) \\
&&\vec{U}_{\sigma}^{A}({\bf r}) =
\vec{U}_\sigma({\bf r}) - C^{\nabla J}\frac{e}{\hbar c}\vec{\nabla}\times [\vec{A}\rho({\bf r})]
-C^{sT}\left ( 2\frac{e}{\hbar c} \vec{A}^T\cdot {\bf J}({\bf r})+\frac{e^2}{\hbar^2 c^2}|\vec{A}|^2\vec{s}({\bf r}) \right ) \\
&&\vec{U}_{\Delta}^{A}({\bf r}) =   \vec{U}_{\Delta}({\bf r})  -C^{j} \frac{e}{\hbar c} \vec{A}\rho({\bf r})\\
&&\vec{\nabla}_{A}\cdot \left  ( B({\bf r}) + \vec{\sigma}\cdot\vec{C}({\bf r}) \right )\vec{\nabla}_{A} =  \left[ \vec{\nabla}_A \left  ( B({\bf r}) + \vec{\sigma}\cdot\vec{C}({\bf r}) \right )\right ]\cdot \vec{\nabla}_A + \nonumber \\
&&\left  ( B({\bf r}) + \vec{\sigma}\cdot\vec{C}({\bf r}) \right )
\left [ \Delta -i\frac{e}{\hbar c} \left ( \vec{A}\cdot \vec{\nabla}_A +\vec{\nabla}_A\cdot \vec{A} \right )  + \frac{e^2}{\hbar^2c^2}|\vec{A}|^2 \right ]
\end{eqnarray}

\begin{center}
{\bf Numerical Implementation}
\end{center}

We build a spatial three-dimensional Cartesian grid in coordinate space with periodic boundary
conditions, and derivatives evaluated in momentum (Fourier-transformed) space. 
This method represents a flexible tool to describe large amplitude nuclear motion as it contains the coupling 
to the continuum and between single-particle and collective degrees of freedom.
For the present problem, we have considered a box  size of $40^3$ with
the lattice constant 1 fm. The time step has been set to $0.076772\ \textrm{fm}/c$ with a
 total time interval of about $4000\ \textrm{fm}/c$. 
The projectile is initially placed at such a distance from the target nucleus that the collision occurs after $1600 - 1700  \ \textrm{fm}/c$. Even though inially the projectle is far enough from the target and hence the EM fields are weak, spurious excitations produced by a sudden switch of the EM interaction at $t=0$ are possible. They were avoided by
multiplying the the EM potentials in Eq. \eqref{LW} by the smoothing function  
$f(t) = 1/\left[ 1 + \exp( (r(t) - R_{0})/a_{0} ) \right]$, where $R_{0}=250$ fm, $a_{0}=50$ fm. This 
ensures that the EM field varies smoothly within the distance $a_{0}$, but  stay approximately
equal to its physical value within the distance $2R_{0}$.

\begin{center}
{\bf Coulomb potential on the lattice}
\end{center}

Here we describe the method used to describe the Coulomb self-interaction of the target nucleus.

Consider the charge distribution $e\rho({\bf r})$:
\begin{eqnarray}
\nabla^{2}\Phi({\bf r}) &=& 4\pi e^{2}\rho({\bf r}) \\
\Phi({\bf r}) &=& \int d^{3}r'\frac{ e^{2}\rho({\bf r}) }{|{\bf r} - {\bf r'}|}
\end{eqnarray}
Note that above we have defined $\Phi$ as $e\Phi$ (note $e^{2}$ in the formula).
After the Fourier transform one gets:
\begin{eqnarray}
\Phi({\bf r}) &=& \int \frac{d^{3}k}{(2\pi)^3} \frac{ e^{2}\rho(\vec{k}) }{k^{2}}\exp (i\vec{k}\cdot{\bf r})
\end{eqnarray}
The above prescription generates however the spurious interaction between neighbouring cells.

Therefore we define the modified potential ($N_{x}, N_{y}, N_{z}$ denote number of equidistant lattice points in each direction,
$L_{i}=N_{i}\Delta x$, $i=x,y,z$, $\Delta x$ is lattice constant):
\begin{eqnarray}
f(r) &=& 1/r \mbox{ for } r<\sqrt{L_{x}^{2}+L_{y}^{2}+L_{z}^{2}} \nonumber \\
f(r) &=& 0   \mbox{ otherwise}
\end{eqnarray}
Clearly the Fourier transform is:
\begin{eqnarray}
f(k) = 4\pi \frac{1-\cos(k\sqrt{L_{x}^{2}+L_{y}^{2}+L_{z}^{2}})}{k^{2}}
\end{eqnarray}
and moreover
\begin{eqnarray}
\Phi({\bf r}) &=& \int \frac{d^{3}k}{(2\pi)^3} \frac{ e^{2}\rho(\vec{k}) }{k^{2}}\exp (i\vec{k}\cdot{\bf r}) 
               = \frac{1}{27 N_{x}N_{y}N_{z}}\sum_{\vec{k} \in L_{x}L_{y}L_{z}} e^{2}\rho(\vec{k}) f(k) \exp (i\vec{k}\cdot{\bf r})
\end{eqnarray}
where in the last term $\rho(\vec{k})$ is the Fourier transformed density on the lattice $27 L_{x}L_{y}L_{z}$.
In practice it means that one has to perform forward and backward Fourier transforms on the lattice three times bigger
in each direction.

This may however be avoided if one realizes that the Fourier transform of the density in the larger lattice can be expressed
through the Fourier transforms in the smaller lattices:
\begin{eqnarray}
\rho_{klm}(\vec{k}) = \sum_{{\bf r} \in L^{3}}\rho(x,y,z)
  \exp\left (-i\left ( k\frac{2\pi}{3L_{x}}x + l\frac{2\pi}{3L_{y}}y + m\frac{2\pi}{3L_{z}}z \right )\right )\exp(-i\vec{k}\cdot{\bf r})
\end{eqnarray}
and we need to perform 27 FFTs on the smaller lattice $L$ for $k,l,m=0,1,2$ of the following quantities: \\
$\rho(x,y,z)\exp\left (-i\left ( k\frac{2\pi}{L_{x}}x + l\frac{2\pi}{L_{y}}y + m\frac{2\pi}{L_{z}}z \right )\right )$

Subsequently we obtain the potential through the relation:
\begin{eqnarray}
& &\Phi({\bf r}) = \nonumber \\
              &=& \frac{1}{27 N_{x}N_{y}N_{z}}\sum_{k,l,m=0}^{2}\left [ \sum_{\vec{k}\in L^{3}} 
e^{2}\rho_{klm}(\vec{k}) f\left (\vec{k} + 
\left ( k\frac{2\pi}{3L_{x}}, l\frac{2\pi}{3L_{y}}, m\frac{2\pi}{3L_{z}} \right ) \right ) \exp (i\vec{k}\cdot{\bf r}) \right ] \nonumber \\
&\times &\exp\left (i\left ( k\frac{2\pi}{3L_{x}}x + l\frac{2\pi}{3L_{y}}y + m\frac{2\pi}{3L_{z}}z \right )\right )
\end{eqnarray}

\begin{center}
{\bf Dipole component of the electromagnetic field produced by the projectile}
\end{center}

\begin{figure}
\centering\includegraphics[scale=0.15]{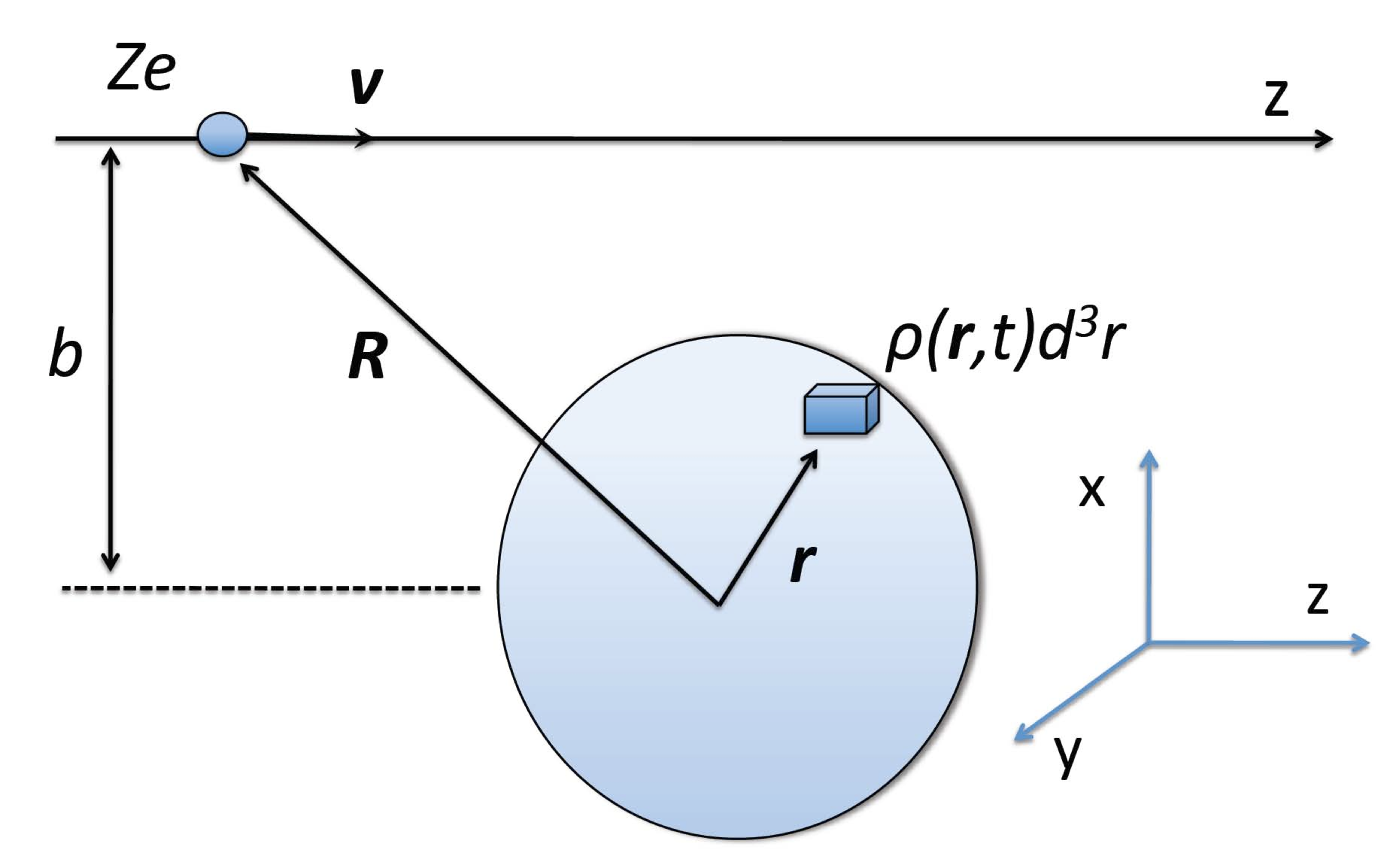}
\caption{Coordinate system used to describe the reaction}\label{fig1} 
\end{figure}

Coordinates (see fig. \ref{fig1}): 
\begin{equation}
{\bf R}= (b,0,vt), \ \ \ \ {\bf r}=(x,y,z).
\end{equation}
Interaction potential:
\begin{equation}
V_E({\bf r},t) = \Phi({\bf r},t) \rho_c({\bf r}) d^3r,\label{VE}
\end{equation}
where  
\begin{equation}
\Phi({\bf r},t)={\gamma Z e \over \sqrt{(x-b)^2+y^2+\gamma^2 (z-vt)^2}},
\end{equation}
and $\gamma = (1-v^2/c^2)^{-1/2}$. $\rho_c({\bf r})=e\Psi^*({\bf r})\Psi({\bf r})$ is the charge density of the nucleus at location ${\bf r}$ and $\Psi({\bf r})$ are proton wavefunctions .
The vector potential is given by
\begin{equation}
{\bf A}({\bf r},t)={{\bf v}\over c}\phi({\bf r},t).
\end{equation}

In order to extract the dipole component we used the interaction Hamiltonian:
\begin{equation}
{\cal H}_{int}({\bf r},t)={\gamma Z e^2 \over \sqrt{(x-b)^2+y^2+\gamma^2 (z-vt)^2}} - {\gamma Z e^2 \over \sqrt{b^2+\gamma^2 v^2t^2}}, \label{hintall}
\end{equation}
where one subtracts the second term which is responsible for the c.m. scattering (i.e. monopole field). 

Consequently the dipole term reads:
\begin{equation}
{\cal H}_{{\rm E1}m}({\bf r},t)
=\sqrt{\frac{2\pi}{3}} r Y_{1m}\left(  \hat{\bf r}\right)  \frac{\gamma Z e^2}{\left(
b^{2}+\gamma ^{2}v^2t^2\right)^{3/2}}\left\{
\begin{array}
[c]{c}%
\mp b,\ \ (\mathrm{if}\ \ \ m=\pm1)\\
\sqrt{2}vt\ \ (\mathrm{if}\ \ \ m=0)\
\end{array}
\right.,  \label{relE1}%
\end{equation}
where {\bf r} is the coordinate of one of the protons in the target.  A sum over $m$ has to be performed, i.e. 
\begin{equation}
{\cal H}_{E1}({\bf r},t)=\sum_{i=protons} \sum_m {\cal H}_{E1m}({\bf r_i},t)=\sum_{i=protons}\sum_m r_i^l Y_{1m}(\hat{\bf r}_i) f_{1m}(t),
\end{equation}
where $f_{1m}(t)$ is the part of the interaction which does not involve the intrinsic structure of the nucleus:
\begin{equation}
f_{1m}(t)=\sqrt{\frac{2\pi}{3}}   \frac{\gamma Z e^2}{\left(
b^{2}+\gamma ^{2}v^2t^2\right)^{3/2}}\left\{
\begin{array}
[c]{c}%
\mp b,\ \ (\mathrm{if}\ \ \ m=\pm1)\\
\sqrt{2}vt\ \ (\mathrm{if}\ \ \ m=0)\
\end{array}
\right. .  \label{fE1t}
\end{equation}

\begin{center}
{\bf Electromagnetic radiation from a nucleus described within TDSLDA}
\end{center}

Let us consider the proton density and current (we use Gauss units):
\begin{eqnarray}
 \rho({\bf r},t) &=& \int_{-\infty}^{\infty} \frac{d\omega}{2\pi} \rho({\bf r},\omega) \exp (-i\omega t) \\
 \vec{j}({\bf r},t) &=& \frac{1}{2\pi}\int_{-\infty}^{\infty} d\omega \vec{j}({\bf r},\omega) \exp (-i\omega t) 
\end{eqnarray}
where
\begin{eqnarray}
\rho({\bf r},\omega) &=& \int \frac{d^{3}k}{(2\pi)^{3}} \rho(\vec{k},\omega) \exp (i\vec{k}\cdot{\bf r})   \\
\vec{j}({\bf r},\omega) &=& \int \frac{d^{3}k}{(2\pi)^{3}} \vec{j}(\vec{k},\omega) \exp (i\vec{k}\cdot{\bf r})  
\end{eqnarray}
Maxwell equations:
\begin{eqnarray}
\nabla\cdot\vec{E}({\bf r},t) &=& 4\pi e\rho({\bf r},t) \\
\nabla\cdot\vec{B}({\bf r},t) &=& 0 \\
\nabla\times\vec{E}({\bf r},t) &=& -\frac{1}{c}\frac{\partial}{\partial t} \vec{B}({\bf r},t) \\
\nabla\times\vec{B}({\bf r},t) &=& \frac{1}{c} \left ( \frac{\partial}{\partial t} \vec{E}({\bf r},t) + 4\pi e\vec{j}({\bf r},t) \right )
\end{eqnarray}
and spatial Fourier transforms:
\begin{eqnarray}
i\vec{k}\cdot\vec{E}(\vec{k},t) &=& 4\pi e\rho(\vec{k},t) \\
i\vec{k}\cdot\vec{B}(\vec{k},t) &=& 0 \\
i\vec{k}\times\vec{E}(\vec{k},t) &=& -\frac{1}{c}\frac{\partial}{\partial t} \vec{B}(\vec{k},t) \\
i\vec{k}\times\vec{B}(\vec{k},t) &=& \frac{1}{c} \left ( \frac{\partial}{\partial t} \vec{E}(\vec{k},t) + 4\pi e\vec{j}(\vec{k},t) \right )
\end{eqnarray}
Hence clearly:
\begin{eqnarray}
\vec{E} &=& \vec{E}_{||} + \vec{E}_{\perp} \\
\vec{B} &=& \vec{B}_{\perp}
\end{eqnarray}
where
\begin{eqnarray}
\nabla\times\vec{E}_{||}({\bf r},t)   &=& 0 \\
\nabla\cdot\vec{E}_{\perp}({\bf r},t) &=& 0 \\
\nabla\cdot\vec{B}_{\perp}({\bf r},t) &=& 0
\end{eqnarray}
and
\begin{eqnarray}
\vec{E} &=& -\nabla\phi({\bf r},t) - \frac{1}{c}\frac{\partial}{\partial t}\vec{A}({\bf r},t)   \\
\vec{B} &=&  \nabla\times\vec{A}({\bf r},t)
\end{eqnarray}
Clearly
\begin{eqnarray}
\vec{E}_{||}({\bf r},t)    &=& -\nabla\phi({\bf r},t)-\frac{1}{c}\frac{\partial}{\partial t}\vec{A}_{||}({\bf r},t)   \\
\vec{E}_{\perp}({\bf r},t) &=& -\frac{1}{c}\frac{\partial}{\partial t}\vec{A}_{\perp}({\bf r},t)   \\
\vec{B}                    &=& \nabla\times\vec{A}_{\perp}({\bf r},t)
\end{eqnarray}
Therefore one has a freedom to choose $\vec{A}_{||}({\bf r},t)$ (gauge transformation) whereas $\vec{A}_{\perp}({\bf r},t)$
is the gauge invariant part of the vector potential.

We choose the Coulomb gauge:
\begin{eqnarray}
\vec{A}_{||}({\bf r},t) = 0 \Leftrightarrow \vec{A}({\bf r},t) = \vec{A}_{\perp}({\bf r},t)
\end{eqnarray}
Hence
\begin{eqnarray}
\vec{E}_{||}({\bf r},t)    &=& -\nabla\phi({\bf r},t)   \\
\vec{E}_{\perp}({\bf r},t) &=& -\frac{1}{c}\frac{\partial}{\partial t}\vec{A}_{\perp}({\bf r},t)   \\
\vec{B}                    &=& \nabla\times\vec{A}_{\perp}({\bf r},t)
\end{eqnarray}
and only perpendicular components of electric and magnetic fields are responsible for emission of radiation.
The important equation in this case is the fourth Maxwell equation:
\begin{eqnarray}
\nabla\times\vec{B}_{\perp}({\bf r},t) &=& \frac{1}{c} \left (  \frac{\partial}{\partial t} \vec{E}_{\perp}({\bf r},t) + 
                                                               4\pi e\vec{j}_{\perp}({\bf r},t) \right ) +
                                          \frac{1}{c} \left (  \frac{\partial}{\partial t} \vec{E}_{||}({\bf r},t) + 
                                                               4\pi e\vec{j}_{||}({\bf r},t) \right )
\end{eqnarray}
Since the lhs represents the vector of type $\perp$ therefore:
\begin{eqnarray}
\frac{\partial}{\partial t} \vec{E}_{||}({\bf r},t) + 4\pi e\vec{j}_{||}({\bf r},t) = 0
\end{eqnarray}
and
\begin{eqnarray}
\nabla\times\vec{B}_{\perp}({\bf r},t) &=& \frac{1}{c} \left (  \frac{\partial}{\partial t} \vec{E}_{\perp}({\bf r},t) + 
                                                               4\pi e\vec{j}_{\perp}({\bf r},t) \right )
\end{eqnarray}
Substituting the potential $\vec{A}$:
\begin{eqnarray}
\nabla\times\nabla\times\vec{A}_{\perp}({\bf r},t) &=& 
-\nabla^{2}\vec{A}_{\perp}({\bf r},t) = \frac{1}{c^{2}} \left (  -\frac{\partial}{\partial t^{2}} \vec{A}_{\perp}({\bf r},t) + 
                                                                  4\pi c e\vec{j}_{\perp}({\bf r},t) \right ) \\
\nabla^{2}\vec{A}_{\perp}({\bf r},t)-\frac{1}{c^{2}}  \frac{\partial}{\partial t^{2}} \vec{A}_{\perp}({\bf r},t) & =  &
-\frac{1}{c}4\pi e\vec{j}_{\perp}({\bf r},t)
\end{eqnarray}
where 
\begin{eqnarray}
\vec{j}_{\perp}({\bf r},t) &=& \vec{j}({\bf r},t) - \vec{j}_{||}({\bf r},t) 
 = \vec{j}({\bf r},t) + \frac{1}{4\pi e}\frac{\partial}{\partial t} \vec{E}_{||}({\bf r},t) 
 = \vec{j}({\bf r},t) - \frac{1}{4\pi e}\frac{\partial}{\partial t} \nabla\phi({\bf r},t)
\end{eqnarray}

Therefore in the Coulomb gauge
\begin{eqnarray}
 \vec{A}({\bf r},t) &=& \vec{A}_{\perp}({\bf r},t) = 
      \frac{1}{c} \int d^{3}r' \frac{e\vec{j}_{\perp}({\bf r'},t-|{\bf r}-{\bf r'}|/c)}{|{\bf r}-{\bf r'}|}
    = \frac{1}{c} \int d^{3}r' \frac{e\vec{j}_{\perp}({\bf r'},\omega)\exp (ik|{\bf r}-{\bf r'}|)}{|{\bf r}-{\bf r'}|}
\end{eqnarray}
and in the far zone $r>>r'$:
\begin{eqnarray}
 \vec{A}({\bf r},\omega) &=& \frac{\exp (ikr)}{r}\frac{1}{c} \int d^{3}r' e\vec{j}_{\perp}({\bf r'},\omega)\exp (-i\vec{k}\cdot{\bf r'})
                          =   \frac{\exp (ikr)}{r}\frac{1}{c} e\vec{j}_{\perp}( \vec{k}, ck)
\end{eqnarray}
where $\vec{k}=k\frac{\bf r}{r}$ and $\omega=c k$ and consequently:
\begin{eqnarray}
 \vec{A}({\bf r},t) &=& \frac{1}{2\pi} \int_{-\infty}^{\infty} d\omega \vec{A}({\bf r},\omega) \exp (-i\omega t)
                     =  \frac{e}{2\pi } \int_{-\infty}^{\infty} dk \vec{j}_{\perp}(\vec{k}, ck) \frac{\exp (ik (r - c t)}{r}
\end{eqnarray}
Consequently since $\vec{B}=\nabla\times\vec{A}$ we get:
\begin{eqnarray}
& &\vec{B}({\bf r},\omega) = \nabla\times\vec{A}({\bf r},\omega)
  =\nabla\times\frac{1}{c} \int d^{3}r' \frac{e\vec{j}_{\perp}({\bf r'},\omega)\exp (ik|{\bf r}-{\bf r'}|)}{|{\bf r}-{\bf r'}|} = \nonumber \\
 &=&\frac{1}{c}\int d^{3}r'\frac{ik ({\bf r}-{\bf r'})\times e\vec{j}_{\perp}({\bf r'},\omega)\exp (ik|{\bf r}-{\bf r'}|)}{|{\bf r}-{\bf r'}|^{2}}  
   -\frac{1}{c}\int d^{3}r' \frac{({\bf r}-{\bf r'})\times e\vec{j}_{\perp}({\bf r'},\omega)\exp (ik|{\bf r}-{\bf r'}|)}{|{\bf r}-{\bf r'}|^{3}} \nonumber \\
 &=&\frac{1}{c}\int d^{3}r'\frac{ik ({\bf r}-{\bf r'})\times e\vec{j}({\bf r'},\omega)\exp (ik|{\bf r}-{\bf r'}|)}{|{\bf r}-{\bf r'}|^{2}}  
   -\frac{1}{c}\int d^{3}r' \frac{({\bf r}-{\bf r'})\times e\vec{j}({\bf r'},\omega)\exp (ik|{\bf r}-{\bf r'}|)}{|{\bf r}-{\bf r'}|^{3}},
\end{eqnarray}
where in the last line we have used the fact that rotation of the vector of type $||$ is zero.
For the electric field:
\begin{eqnarray}
\vec{E}_{\perp}({\bf r},\omega) &=& \frac{i\omega}{c}\vec{A}({\bf r},\omega)
= \frac{i\omega}{c^{2}} \int d^{3}r' \frac{e\vec{j}_{\perp}({\bf r'},\omega)\exp (ik|{\bf r}-{\bf r'}|)}{|{\bf r}-{\bf r'}|} \\
\vec{E}_{||}({\bf r},\omega) &=& -\nabla\phi({\bf r},\omega) = -\nabla \int d^{3}r'\frac{e \rho({\bf r'},\omega)}{|{\bf r}-{\bf r'}|}
\end{eqnarray}

Hence in the far zone $r>>r'$ one gets:
\begin{eqnarray}
\vec{B}({\bf r},\omega) &=& \frac{ie}{c}\frac{\exp(ikr)}{r}\int d^{3}r'\vec{k}\times \vec{j}_{\perp}({\bf r'},\omega)
\exp(-i\vec{k}\cdot{\bf r'}) = \frac{ie}{c}\frac{\exp (ikr)}{r}\vec{k}\times\vec{j}(\vec{k},\omega)  \\
\vec{E}({\bf r},\omega) &=& \vec{E}_{\perp}({\bf r},\omega) + \vec{E}_{||}({\bf r},\omega) 
                                 = \frac{\exp (ikr)}{r}\frac{i\omega}{c^{2}} e\vec{j}_{\perp}( \vec{k}, ck) 
                                 =  \frac{ie}{c}\frac{\exp (ikr)}{r} k \vec{j}_{\perp}( \vec{k}, ck)  \\
                                &=& \frac{ie}{c}\frac{\exp (ikr)}{r} k ( \vec{j}( \vec{k}, ck) - \vec{j}_{||}( \vec{k}, ck) )
                                 =  \frac{ie}{c}\frac{\exp (ikr)}{r} 
               \left ( ( \vec{k}\cdot\frac{\bf r}{r} )\vec{j}( \vec{k}, ck) - ( \vec{k}\cdot\vec{j}( \vec{k}, ck) )\frac{\bf r}{r} \right ) \\
               &=&  \frac{ie}{c}\frac{\exp (ikr)}{r} \frac{\bf r}{r}\times\left ( \vec{j}(\vec{k},\omega) \times\vec{k} \right )
\end{eqnarray}

and consequently:
\begin{eqnarray}
 \vec{B}({\bf r},t) = \frac{1}{2\pi} \int_{-\infty}^{\infty} d\omega \vec{B}({\bf r},\omega) \exp (-i\omega t) &=& 
      \frac{ie}{2\pi}\int_{-\infty}^{\infty} dk \vec{k}\times\vec{j}(\vec{k},ck)\frac{\exp (ik (r - c t) )}{r} \\
 &=&  \frac{ie}{c}\int_{-\infty}^{\infty} \frac{d\omega}{2\pi} \vec{k}\times\vec{j}(\vec{k},\omega)\frac{\exp (-i\omega (t - r/c) )}{r} \\
 \vec{E}({\bf r},t) = \frac{1}{2\pi} \int_{-\infty}^{\infty} d\omega \vec{E}({\bf r},\omega) \exp (-i\omega t) &=&
 \frac{ie}{2\pi}\frac{\bf r}{r}\times \int_{-\infty}^{\infty} dk \left ( \vec{j}(\vec{k},ck) \times \vec{k} \right ) 
 \frac{\exp (ik (r - c t) )}{r} \\
&=& \frac{ie}{c}\frac{\bf r}{r}\times \int_{-\infty}^{\infty} \frac{d\omega}{2\pi} \left ( \vec{j}(\vec{k},\omega) \times \vec{k} \right ) 
 \frac{\exp ( -i\omega ( t - r/c ) )}{r} 
\end{eqnarray}

Note that in the above expressions $\vec{k}$ and $\omega$ are related: $\omega=c|\vec{k}|$. 

Poynting vector reads $\vec{S}=\frac{c}{4\pi}\vec{E}\times\vec{B}$ and thus:
\begin{eqnarray}
\vec{S}(t) &=& \frac{c}{4\pi (2\pi)^2}\int_{-\infty}^{\infty} d\omega\int_{-\infty}^{\infty} d\omega' \vec{E}({\bf r},\omega)\times \vec{B}({\bf r},\omega)\exp (-i(\omega+\omega')t)
\nonumber \\
 &=& \frac{c}{4\pi (2\pi)^2}\int_{-\infty}^{\infty} d\omega\int_{-\infty}^{\infty} d\omega' \vec{E}({\bf r},\omega)\times \vec{B}^{*}({\bf r},\omega')\exp (-i(\omega-\omega')t) 
 \nonumber \\
 &=& \frac{e^{2}}{4\pi (2\pi)^2 c r^{2}}\frac{\bf r}{r}\int_{-\infty}^{\infty} d\omega\int_{-\infty}^{\infty} d\omega' 
 ( \vec{k}\times \vec{j}(\vec{k},\omega) )\cdot( \vec{k'}\times\vec{j}^{*}(\vec{k'},\omega') ) \exp (-i(\omega-\omega')t+i(k-k')r) \\
  &=& \frac{e^{2}}{4\pi c r^{2}}\frac{\bf r}{r}\left |\int_{-\infty}^{\infty} \frac{d\omega}{2\pi}
 ( \vec{k}\times \vec{j}(\vec{k},\omega) )\exp (-i \omega t+ikr)\right |^{2} \\
 &=& \frac{c}{4\pi }\frac{\bf r}{r}\left |\int_{-\infty}^{\infty} \frac{d\omega}{2\pi}
 \vec{B}({\bf r},\omega) \exp (-i \omega t) \right |^{2}
\end{eqnarray}

Energy per unit time emitted to the angle $d\Omega$ reads:
\begin{eqnarray}
dP(t) &=& \vec{S}(t)\cdot\frac{\bf r}{r}r^{2} d\Omega = \frac{e^{2}}{4\pi c }
                     \left | \int \frac{d\omega}{2\pi}
                          ( \vec{k}\times \vec{j}(\vec{k},\omega) )\exp (-i \omega t+ikr) \right |^{2}d\Omega  
\end{eqnarray}
Hence
\begin{eqnarray} \label{powert}
\frac{dP}{d\Omega}(t) &=& \frac{e^{2}}{4\pi c }\left |\int_{-\infty}^{\infty} \frac{d\omega}{2\pi}
                          ( \vec{k}\times \vec{j}(\vec{k},\omega) )\exp (-i \omega ( t - r/c ) ) \right |^{2} 
                      = \frac{c}{4\pi}r^{2}\left | \vec{B}({\bf r},t)\right |^{2}
\end{eqnarray}
Note that the radiation at time $t$ is given by the current at time $t-r/c$, thus a simple time shift, which we can discard. 
Therefore the total amount of radiated energy at the angle $d\Omega$ reads:
\begin{eqnarray}
\int_{-\infty}^{\infty}\frac{dP}{d\Omega}(t)dt 
          &=& \frac{c}{4\pi}r^{2}\int_{-\infty}^{\infty} \left | \vec{B}({\bf r},t)\right |^{2} dt =
            \frac{c}{4\pi}r^{2} \int_{-\infty}^{\infty} \vec{B}({\bf r},t)\cdot
            \int_{-\infty}^{\infty}\frac{d\omega}{2\pi} \vec{B}({\bf r},\omega)\exp (-i \omega t) dt \nonumber \\
        &=& \frac{c}{4\pi}r^{2} 
            \int_{-\infty}^{\infty}\frac{d\omega}{2\pi} \vec{B}({\bf r},-\omega)\cdot\vec{B}({\bf r},\omega) =
            \frac{c}{4\pi}r^{2} 
            \int_{-\infty}^{\infty}\frac{d\omega}{2\pi} \vec{B}^{*}({\bf r},\omega)\cdot\vec{B}({\bf r},\omega) 
\end{eqnarray}
which gives the spectral decomposition of emitted radiation:
\begin{eqnarray}
\int_{-\infty}^{\infty}\frac{dP}{d\Omega}(t)dt 
          &=& 
  \frac{c}{8\pi^2}r^{2} \int_{-\infty}^{\infty}d\omega \left | \vec{B}({\bf r},\omega) \right |^{2} 
  =   \frac{c}{4\pi^2}r^{2} \int_{0}^{\infty}d\omega \left | \vec{B}({\bf r},\omega) \right |^{2} 
\end{eqnarray}
Hence the energy emitted at the angle $\Omega$ at frequency $\omega$  reads:
\begin{eqnarray} \label{spectr}
\frac{dE}{d\Omega d\omega}(\omega) 
          &=& \frac{e^2}{4\pi^2 c} \left | \vec{k}\times \vec{j}(\vec{k},\omega) \right |^{2} 
           =  \frac{e^2}{4\pi^2 c} \left | \int d^{3}r 
              \left ( \nabla \times \vec{j}({\bf r},\omega)  \right )
              \exp(-i\vec{k}\cdot{\bf r}) \right |^{2} 
\end{eqnarray}

In order to calculate the quantities given by the expressions: (\ref{powert}) (\ref{spectr})
we use the multipole expansion.
Namely, let us consider eq. (\ref{spectr}):
\begin{eqnarray}
\frac{dE}{d\omega} &=& \int\frac{dE}{d\Omega d\omega}(\omega) d\Omega 
          =  \int \frac{e^2}{4\pi^2 c} \left | \vec{k}\times \vec{j}(\vec{k},\omega) \right |^{2}d\Omega \\
          &=&  \frac{e^2}{4\pi^2 c} \int \left | \int d^{3}r \int_{-\infty}^{\infty} dt \left ( \vec{\nabla}\times \vec{j}({\bf r},t) \right ) 
                                         \exp(-i\vec{k}\cdot {\bf r} + i\omega t) \right |^{2}d\Omega 
\end{eqnarray}
Let us denote:
\begin{eqnarray}
\vec{\nabla}\times \vec{j}({\bf r},t) &=& \vec{b}({\bf r},t) \\
\vec{\nabla}\times \vec{j}({\bf r},\omega) &=& \vec{b}({\bf r},\omega)
\end{eqnarray}
We expand $\exp(-i\vec{k}\cdot {\bf r}) $:
\begin{eqnarray}
\exp(-i\vec{k}\cdot {\bf r}) = 4\pi \sum_{l,m} (-i)^{l} j_{l}(kr)Y_{lm}(\hat{k})Y_{lm}^{*}(\hat{r})
\end{eqnarray}
and consequently we get 
\begin{eqnarray}
\frac{dE}{d\omega} 
          &=&  \frac{e^2}{4\pi^2 c} 
               \int \left | \int d^{3}r \int_{-\infty}^{\infty} dt  \left ( \vec{b}({\bf r},t)
                           4\pi \sum_{l,m} (-i)^{l} j_{l}(kr)Y_{lm}(\hat{k})Y_{lm}^{*}(\hat{r}) \right )
                          \exp(i\omega t) \right |^{2}d\Omega \\
          &=&   \frac{e^2}{4\pi^2 c} 
               \int \left | \int_{-\infty}^{\infty} dt 4\pi \sum_{l,m} (-i)^{l} \vec{b}_{lm}(\vec{k},t)Y_{lm}(\hat{k})
                          \exp(i\omega t) \right |^{2}d\Omega \\
          &=&   \frac{e^2}{4\pi^2 c} 
               \int \left | 4\pi \sum_{l,m} (-i)^{l} \vec{b}_{lm}(\vec{k},\omega)Y_{lm}(\hat{k}) \right |^{2}d\Omega 
\end{eqnarray}
where
\begin{eqnarray}
\vec{b}_{lm}(k,t)      &=& \int d^{3}r\vec{b}({\bf r},t)j_{l}(kr)Y_{lm}^{*}(\hat{r}) \\
\vec{b}_{lm}(k,\omega) &=& \int_{-\infty}^{\infty} \vec{b}_{lm}(k,t)\exp(i\omega t) dt
\end{eqnarray}
Note that $\vec{b}_{lm}$ is a function of $k$ (not $\vec{k}$) and
\begin{eqnarray}
\frac{dE}{d\omega}  &=&  \frac{e^2}{4\pi^2 c} (4 \pi )^{2} 
\int \left ( \sum_{l,m, l', m'} (-i)^{l}i^{l'} \left ( \vec{b}_{lm}(k,\omega)\cdot\vec{b}_{l'm'}^{*}(k,\omega) \right )
                                 Y_{lm}(\hat{k}) Y_{l'm'}^{*}(\hat{k}) \right )d\Omega \\
                    &=&  \frac{e^2}{4\pi^2 c} (4 \pi )^{2} \sum_{l,m} |\vec{b}_{lm}(k,\omega)|^{2} 
                     =   \frac{4 e^2}{c} \sum_{l,m} |\vec{b}_{lm}(k,\omega)|^{2} 
\end{eqnarray}
The above equation is used to calculate the spectrum of emitted radiation. In practice one needs
only few multipoles. The contribution coming from $l=4$ term is already negligibly small.

In order to determine the rate of emitted radiation let us consider eq. (\ref{powert}):
\begin{eqnarray} 
P(t + r/c) &=& \int \frac{dP}{d\Omega}(t + r/c)d\Omega = \frac{e^{2}}{4\pi c }\int \left |\int_{-\infty}^{\infty} \frac{d\omega}{2\pi}
 ( \vec{k}\times \vec{j}(\vec{k},\omega) )\exp (-i \omega t) \right |^{2} d\Omega \\
     &=& \frac{e^{2}}{4\pi c }\int \left |\int_{-\infty}^{\infty} \frac{d\omega}{2\pi}
 \left ( \int d^{3} r ( \vec{\nabla}\times \vec{j}({\bf r},\omega) )\exp(-i\vec{k}\cdot {\bf r}) \right ) 
      \exp (-i \omega t) \right |^{2} d\Omega \\
     &=& \frac{e^{2}}{4\pi c }\int \left |\int_{-\infty}^{\infty} \frac{d\omega}{2\pi}
 \left ( \int d^{3} r 4\pi \sum_{l,m} (-i)^{l}  \vec{b}({\bf r},\omega) j_{l}(kr)Y_{lm}(\hat{k})Y_{lm}^{*}(\hat{r}) \right ) 
      \exp (-i \omega t) \right |^{2} d\Omega \\
     &=& \frac{e^{2}}{4\pi c }\int \left |\int_{-\infty}^{\infty} \frac{d\omega}{2\pi}
  \left (  4\pi \sum_{l,m} (-i)^{l} \vec{b}_{lm}(k,\omega)Y_{lm}(\hat{k})\right ) 
     \exp (-i \omega t) \right |^{2} d\Omega \\
     &=& \frac{e^{2}}{4\pi c } \int_{-\infty}^{\infty} \frac{d\omega}{2\pi}\int_{-\infty}^{\infty} \frac{d\omega'}{2\pi}
\left ( (4\pi)^{2}  \int d\Omega \sum_{l,m,l',m'} (-i)^{l} i^{l'} \left ( \vec{b}_{lm}(k,\omega)\cdot \vec{b}_{l'm'}^{*}(k',\omega') \right )
                                       Y_{lm}(\hat{k})Y_{l'm'}^{*}(\hat{k})\right ) \times \nonumber \\
     &\times& \exp (-i (\omega - \omega ') t)  
\end{eqnarray}
Note that in the last two lines of the above expression $\hat{k}=\hat{k'}$ because the vectors $\vec{k}, \vec{k}'$ differ only
by length ($\omega = ck, \omega'=ck'$) but have the same direction specified by the angle $\Omega$. Therefore:
\begin{eqnarray} 
P(t + r/c) &=&  \frac{e^{2}}{4\pi c } \int_{-\infty}^{\infty} \frac{d\omega}{2\pi}\int_{-\infty}^{\infty} \frac{d\omega'}{2\pi}
 \left ( (4\pi)^{2} \sum_{l,m} \left ( \vec{b}_{lm}(k,\omega)\cdot \vec{b}_{lm}^{*}(k',\omega') \right )\right )\exp (-i (\omega - \omega ') t) \\
  &=&  \frac{e^{2}}{\pi c } \sum_{l,m} \left | \int_{-\infty}^{\infty} \vec{b}_{lm}(k,\omega) \exp (-i \omega t ) d\omega \right |^{2}
\end{eqnarray}
The last equation is used in practice to calculate the rate of emitted radiation.

The above prescriptions work efficiently if one considers the radiation emitted due to internal nuclear excitation.
However in order to determine the contribution coming from the CM motion of the nucleus the simpler
formula can be derived.
In this case the proton current reads:
\begin{eqnarray}
\vec{j}_{p}({\bf r}, t) = \vec{V}(t)\delta( {\bf r} - {\bf r}_{0}(t) )
\end{eqnarray} 
Then
\begin{eqnarray}
\vec{A}({\bf r},\omega) &=& \frac{\exp (ikr)}{r}\frac{Ze}{c} \int d^{3}r' \vec{j}_{p}({\bf r'},\omega)\exp (-i\vec{k}\cdot{\bf r'}) \\
&=& \frac{\exp (ikr)}{r}\frac{1}{c} Ze \int_{-\infty}^{\infty} dt \exp(i\omega t) \vec{V}(t) \exp(-i\vec{k} \cdot {\bf r}_{0}(t) )
\end{eqnarray}
where 
\begin{eqnarray}
\vec{j}_{p}(\vec{k}, \omega) &=& 
\int_{-\infty}^{\infty} dt \exp\left ( i\omega \left ( t - \frac{1}{c}\vec{n}\cdot {\bf r}_{0}(t) \right ) \right ) \vec{V}(t)  \\
 &\approx& \int_{-\infty}^{\infty} dt \exp(i\omega t) \vec{V}\left (t + \frac{1}{c}\vec{n}\cdot {\bf r}_{0}(t) \right )  \\
 &\approx&  \int_{-\infty}^{\infty} dt \exp(i\omega t) \vec{V}(t) = \vec{V}(\omega),
\end{eqnarray}
where $\omega = ck$.
The approximation was made above that the velocity is small and the movement of the nucleus is negligible. Therefore
the possible perturbation of the radiation due to the change of nucleus position can be neglected.

Consequently:
\begin{eqnarray}
\vec{B}({\bf r},\omega) = \frac{i Ze}{c}\frac{\exp (ikr)}{r}\vec{k}\times\vec{V}(\omega) 
\end{eqnarray}
and
\begin{eqnarray}
\vec{B}({\bf r},t)&=&\int_{-\infty}^{\infty}\frac{d\omega}{2\pi}\vec{B}({\bf r},\omega)\exp(-i\omega t) \\
&=&\frac{i Ze}{c^2}\frac{1}{r}\int_{-\infty}^{\infty}\frac{d\omega}{2\pi}
   \exp\left (-i\omega\left (t-\frac{r}{c}\right )\right )\omega\vec{n}\times\vec{V}(\omega) \\
&=&-\frac{Ze}{c^2}\frac{1}{r}\frac{d}{dt}\int_{-\infty}^{\infty}\frac{d\omega}{2\pi}
   \exp\left (-i\omega\left (t-\frac{r}{c}\right )\right )\vec{n}\times\vec{V}(\omega) \\
&=&-\frac{Ze}{c^2}\frac{1}{r}\vec{n}\times\frac{d}{dt}\vec{V}\left (t-\frac{r}{c} \right )
\end{eqnarray}
Therefore 
\begin{eqnarray}
\frac{dP}{d\Omega}(t) &=& \frac{c}{4\pi}r^{2}\left | \vec{B}({\bf r},t)\right |^{2} 
= \frac{1}{4\pi c^{3}} (Ze)^{2} \left |\frac{d V\left (t-\frac{r}{c} \right )}{dt} \right |^{2} \sin^{2}\theta
\end{eqnarray}
and 
\begin{eqnarray} \label{cmpower}
P(t) &=& \frac{2}{3} \frac{(Ze)^{2}}{c^{3}} \left |\frac{d V\left (t-\frac{r}{c} \right )}{dt} \right |^{2} 
\end{eqnarray}

Spectral decomposition:
\begin{eqnarray} 
\frac{dE}{d\Omega d\omega}=\frac{(Ze)^{2}}{4\pi^{2}c} |\vec{k}\times\vec{V}(\omega)|^{2}
\end{eqnarray}
and integrating over angles
\begin{eqnarray} \label{cmspectr}
\frac{dE}{d\omega} = \frac{2}{3\pi}\frac{(Ze)^{2}}{c}k^{2}|\vec{V}(\omega)|^{2} = \frac{2}{3\pi}\frac{(Ze)^{2}}{c^{3}}\omega^{2}|\vec{V}(\omega)|^{2}
= \frac{2}{3\pi}\frac{(Ze)^{2}}{c^{3}}\left |\frac{d\vec{V}}{dt}(\omega) \right |^{2}
\end{eqnarray}
where $\frac{d\vec{V}}{dt}(\omega)$ is the Fourier transform of acceleration:
\begin{eqnarray}
\frac{d\vec{V}}{dt}(\omega) = \int dt \frac{d\vec{V}}{dt} (t) \exp (i\omega t)
\end{eqnarray}

The above derivation assumes that the moving nucleus can be treated as a point-like particle.
This is a reasonable approximation although it is not difficult to include suitable corrections. 
Let us consider the proton current in the form:
\begin{eqnarray}
\vec{j}_{p}({\bf r}, t) = \vec{V}(t)\rho( {\bf r} - {\bf r}_{0}(t) )
\end{eqnarray}
Using the same assumption as before, ie. that the motion is nonrelativistic and
movement in space is negligible one gets:
\begin{eqnarray}
\vec{j}_{p}(\vec{k}, \omega) &=& \vec{V}(\omega)\rho(\vec{k})
\end{eqnarray}
and (see (\ref{spectr})):
\begin{eqnarray}
\int\frac{dE}{d\Omega d\omega}d\Omega &=& 
\frac{(Ze)^{2}}{4\pi^{2}c}| \vec{k}\times \vec{V}(\omega)\rho(\vec{k}) |^{2}d\Omega = 
\frac{(Ze)^{2}}{4\pi^{2}c}\frac{1}{c^{2}} 
\int |\vec{n}\times\omega\vec{V}(\omega)\rho(\vec{k})|^{2} d\Omega \\
&=& \frac{(Ze)^{2}}{4\pi^{2} c^{3}}
\int \left |\vec{n}\times \frac{d\vec{V}}{dt}(\omega) \rho(\vec{k}) \right |^{2} d\Omega
\end{eqnarray}
where $\vec{n} = \frac{\bf r}{r}$. In the case of spherical density distribution it
simplifies to:
\begin{eqnarray}
\int\frac{dE}{d\Omega d\omega}d\Omega &=& \frac{(Ze)^{2}}{4\pi^{2} c^{3}}
\int \left |\vec{n}\times \frac{d\vec{V}}{dt}(\omega)\right |^{2} 
|\rho(k)|^{2} d\Omega = \frac{2}{3\pi}\frac{(Ze)^{2}}{c^{3}}
\left |\frac{d\vec{V}}{dt}(\omega)\right |^{2} |\rho(k)|^{2}  
\end{eqnarray}

\begin{figure}[htb] 
\includegraphics[width=9.0cm]{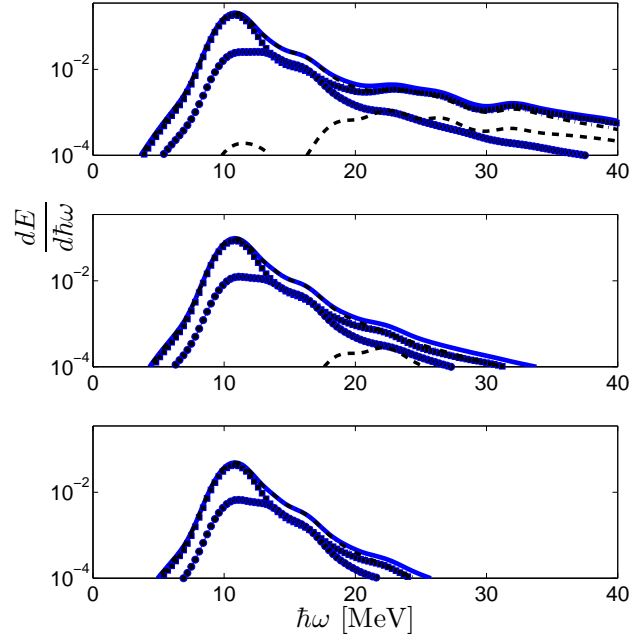}
\caption{ (color online) 
The energy spectrum of emitted electromagnetic radiation due to internal
excitation of the target nucleus, caused by the collision at the impact parameters 
$b=12.2$ fm (upper subfigure), $b=16.2$ and $b=20.2$ (lowest subfigure) .
The contributions from two orientations of the target nucleus are shown: perpendicular (squares)
and parallel (circles) with respect to the incoming projectile.
Dotted dashed line represents the dipole component of the radiation.
Dashed line represents the quadrupole component of the radiation. 
In this case the smoothing width of the original curves was set to $1$ MeV.
}\label{fig2s}
\end{figure}

\begin{figure}[htb] 
\includegraphics[width=9.0cm]{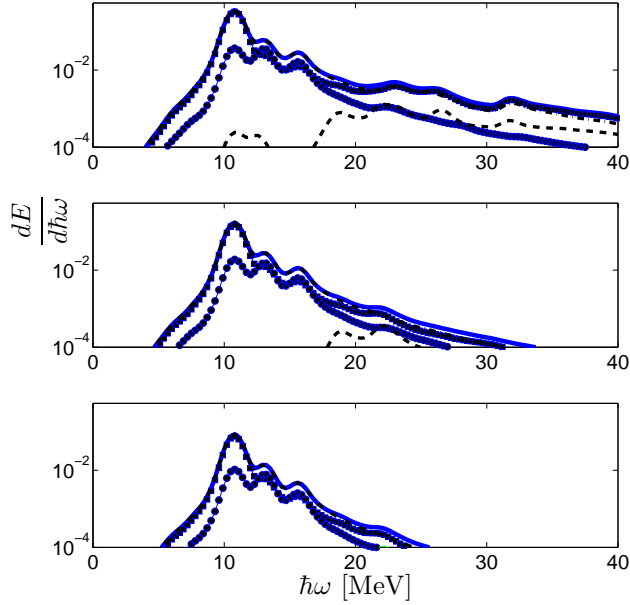}
\caption{ (color online) 
The same as in the Fig. \ref{fig2s}, but in this case the smoothing width of the original curves was set to $0.5$ MeV.
}\label{fig3s}
\end{figure}

The above expressions can be used to determine the spectrum of emitted radiation. In the Figures
\ref{fig2s} and \ref{fig3s} the contributions to the energy spectrum coming from dipole and quadrupole
terms are plotted for $3$ values of impact parameter. The difference between the figures 
originates from two different smoothing widths that have been applied.
Namely, the original curves have been convoluted with gaussians of widths $1$ MeV (Fig. \ref{fig2s}) 
and $0.5$ MeV (Fig. \ref{fig3s} ).

\begin{figure}[htb]
\includegraphics[width=8.0cm]{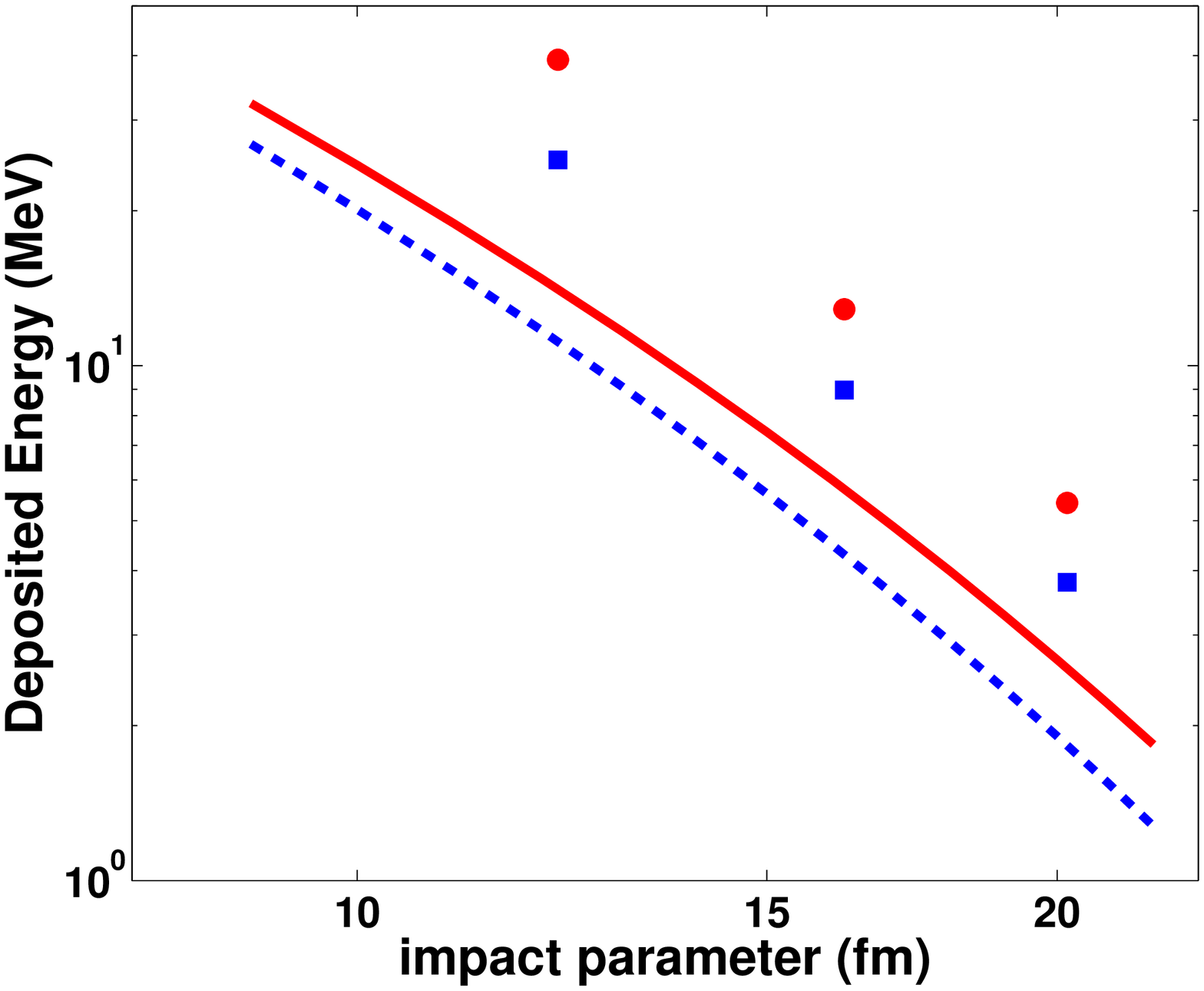}
\caption{ (color online) Energy deposited in the target nucleus $^{238}$U for three values of the impact parameter: $12.2, 16.2, 20.2$ fm and for two nuclear orientations: nuclear symmetry axis being parallel 
(squares) and perpendicular (circles) to the trajectory of incoming projectile. 
The same quantity is shown for the Goldhaber-Teller model, assuming that
the frequencies of the dipole oscillations are $\hbar\omega=12$ MeV and $\hbar\omega=16$ MeV parallel (blue-dashed line)
and perpendicular (red-solid line) to the nuclear symmetry axis, respectively.   }\label{fig1}
\end{figure}

\begin{figure} 
\includegraphics[width=9.0cm]{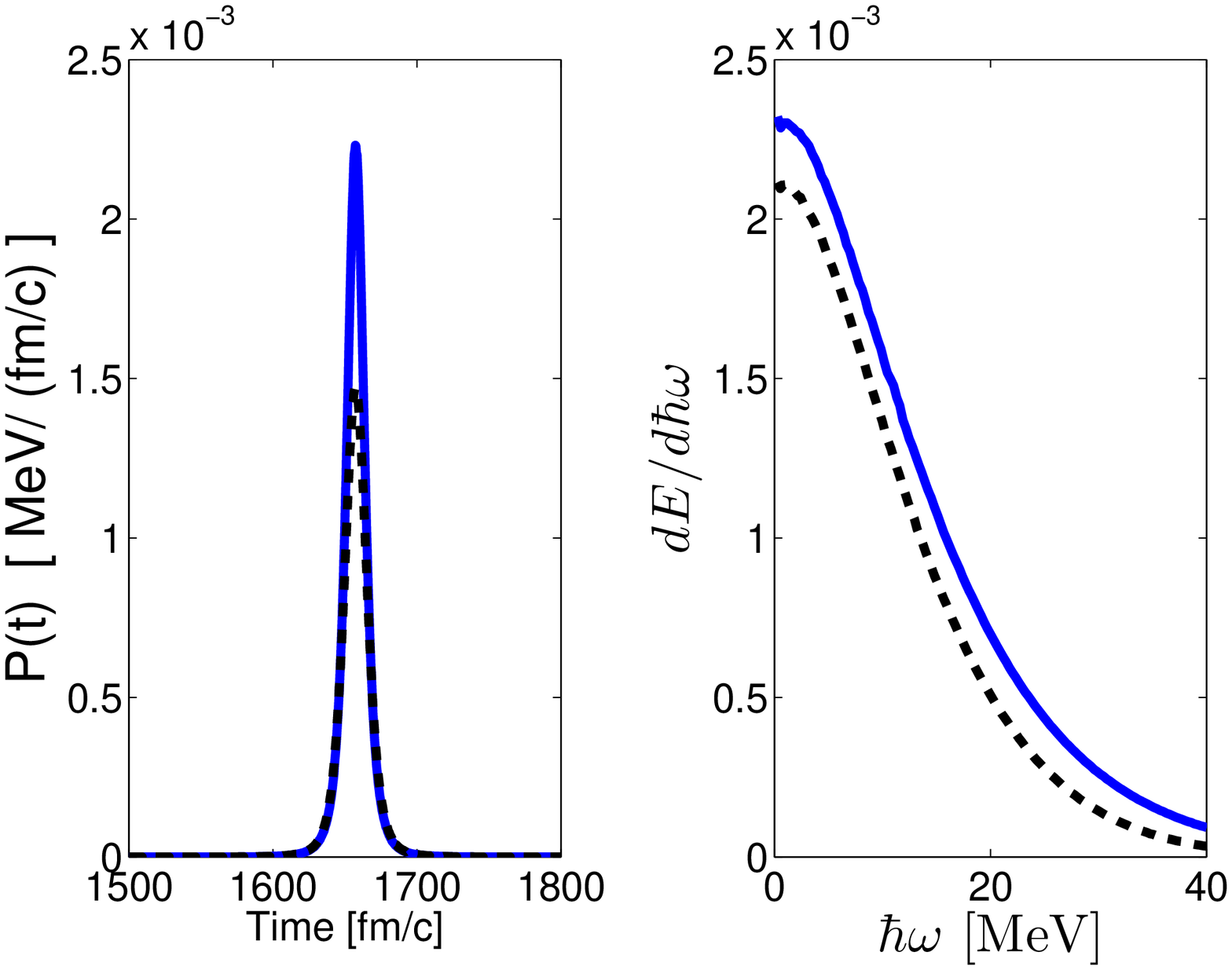}
\caption{ (color online) 
The gamma emission rate (left panel) due to bremsstrahlung for the collision at the
impact parameter $b=12.2$fm. The right panel shows
the energy spectrum emitted. Solid and dashed lines correspond to the perpendicular and parallel orientation
of the target nucleus with respect to incoming projectile.
}\label{fig4s}
\end{figure}

\begin{figure} 
\includegraphics[width=9.0cm]{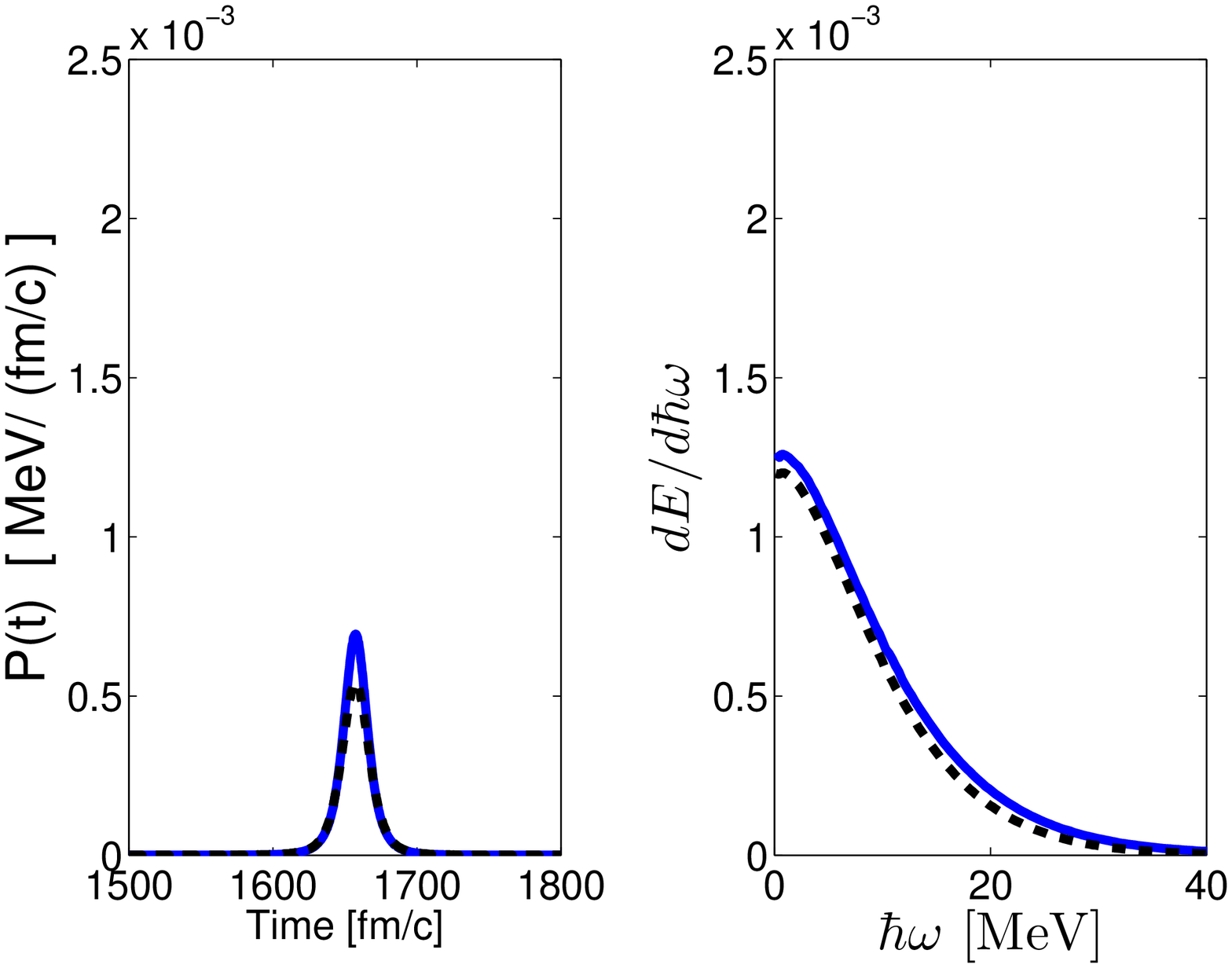}
\caption{ (color online) 
The same as in the Fig. \ref{fig4s}, but for the impact parameter $b=16.2$fm.
}\label{fig5s}
\end{figure}

\begin{figure} 
\includegraphics[width=9.0cm]{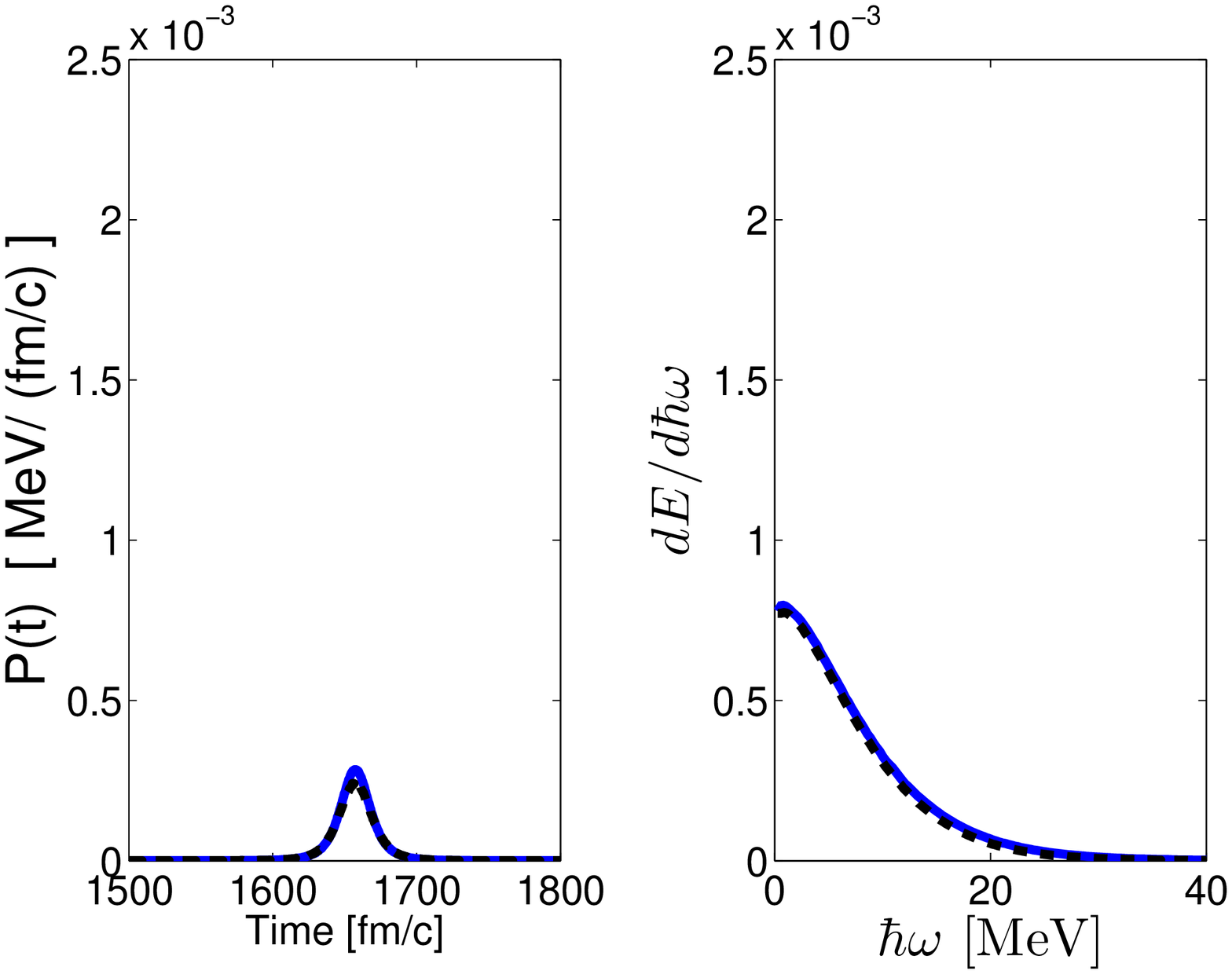}
\caption{ (color online) 
The same as in the Fig. \ref{fig4s}, but for the impact parameter $b=20.2$fm.
}\label{fig6s}
\end{figure}

For the radiation caused by the CM acceleration after collision the decomposition into multipoles is
not useful and one can apply instead eqs. (\ref{cmpower,cmspectr}). The emission occurs within much shorter time
scale governed by the collision time $\tau_{coll}=\frac{b}{v\gamma}$. The results are plotted in 
the Figs. \ref{fig4s}, \ref{fig5s}, \ref{fig6s}.

\begin{figure}[htb] 
\includegraphics[width=8.0cm]{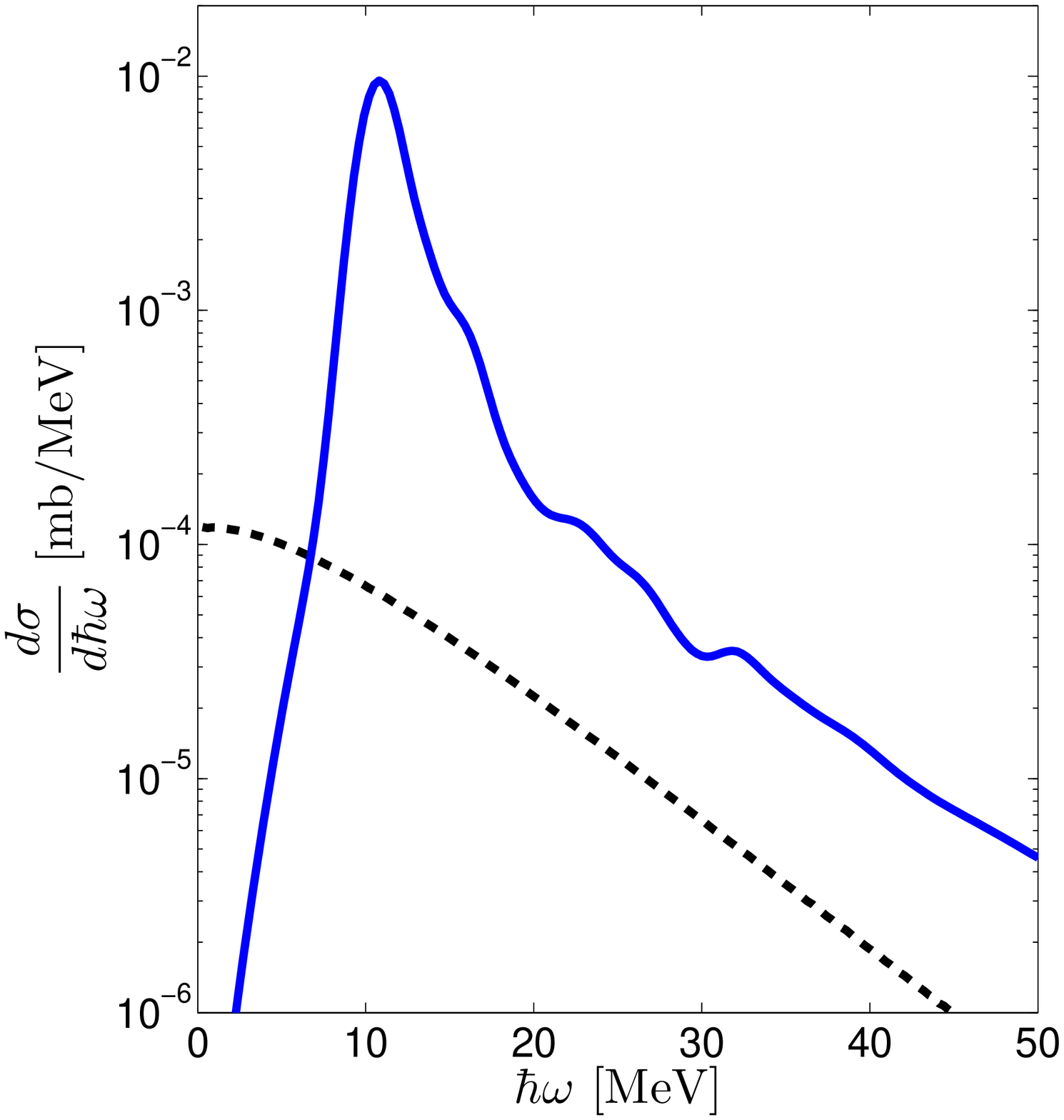}
\caption{ (color online) 
The contributions to the cross section with respect to gamma emission during 
$2500$ fm/c after collision. The dashed line represents the Bremsstrahlung contribution. The solid line shows the contribution from intrinsic excitation
modes.  
}\label{fig5}
\end{figure}

\begin{center}
{\bf Dipole dynamics and neutron emission}
\end{center}

\begin{figure}[htb] 
\includegraphics[width=8.0cm]{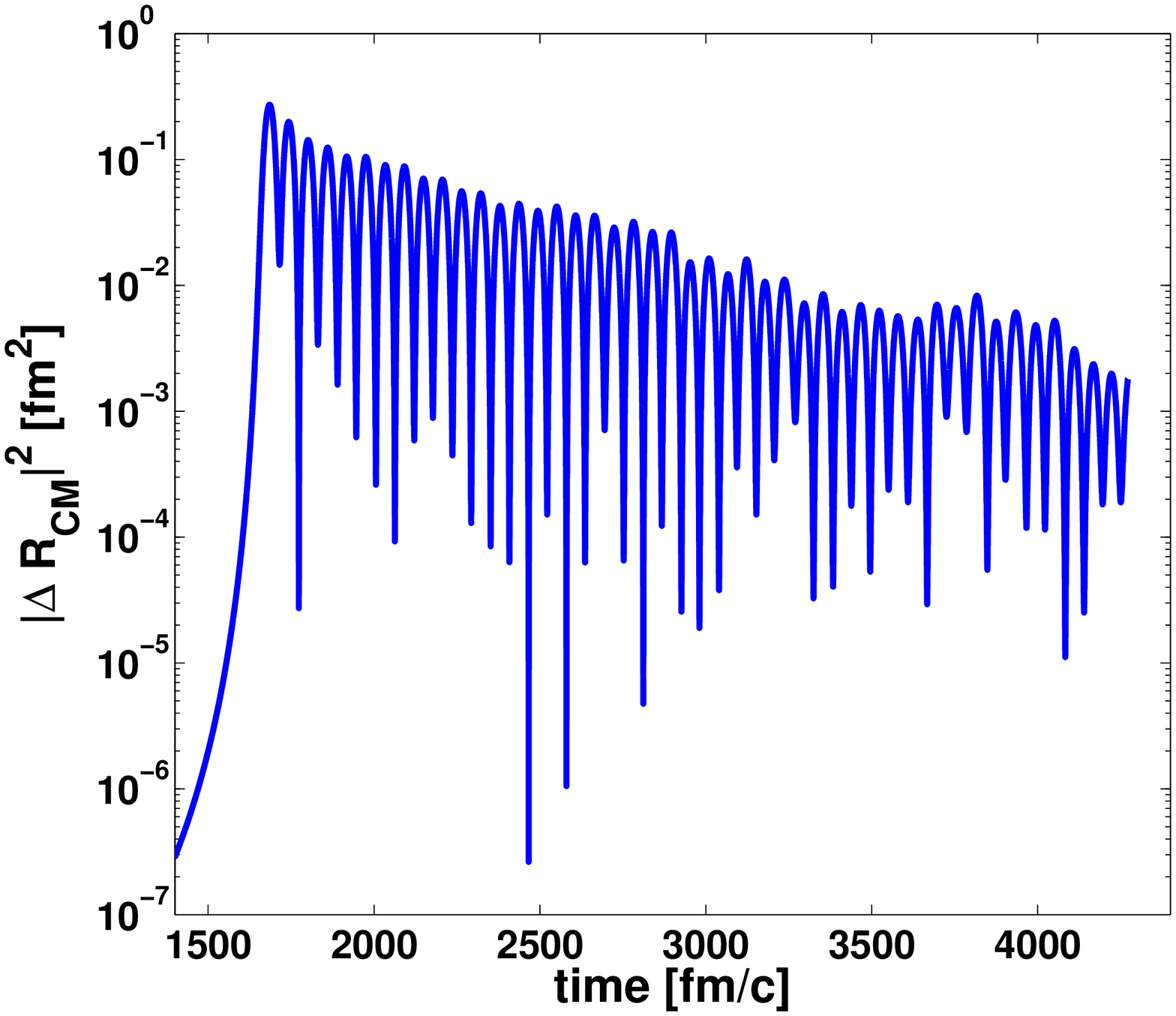}
\caption{ (color online) 
The distance squared between the CM of protons and the total nuclear CM: $|\Delta R_{CM}|^{2}$
as a function of time. The impact parameter is $b=12.2$ fm, and the nuclear symmetry
axis of the target is perpendicular to the projectile's trajectory. 
The slope does not depend on the orientation and the impact parameter. 
The numerical fit to the maxima (squared amplitudes of dipole oscillations)
with the function $\exp(-t/\tau)$ yields $\tau\approx 500$ fm/c.}\label{fig2}
\end{figure}

\begin{figure} 
\includegraphics[width=9.0cm]{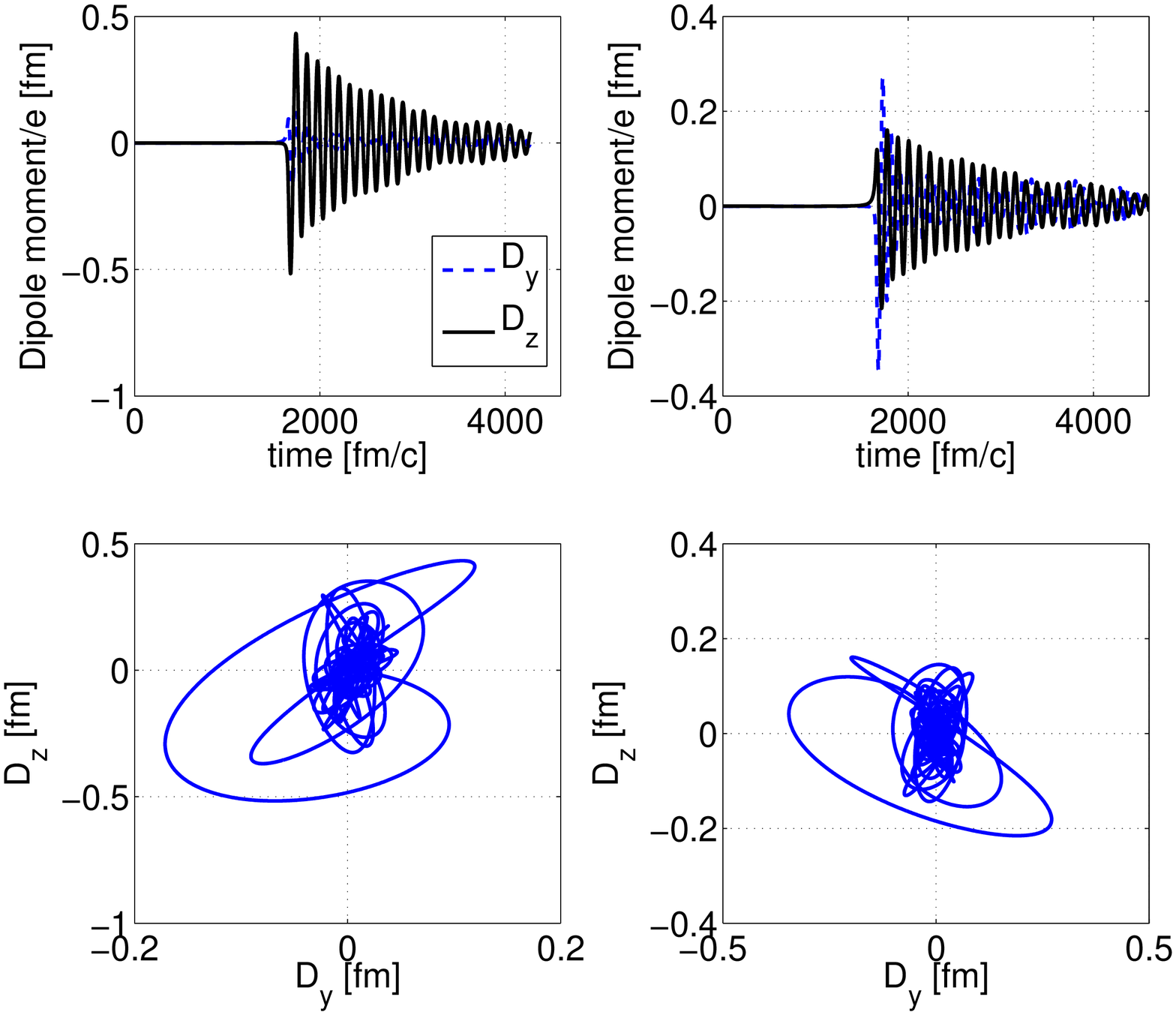}
\caption{ (color online) 
Two components of the dipole moment: $D_{z}$  and $D_{y}$ as a function of time.
The left and right subfigures correspond to the collision with projectile moving along the 
$y$-axis and $z$-axis , respectively. Impact parameter: $b=12.2$fm. 
}\label{fig7s}
\end{figure}

\begin{figure} 
\includegraphics[width=9.0cm]{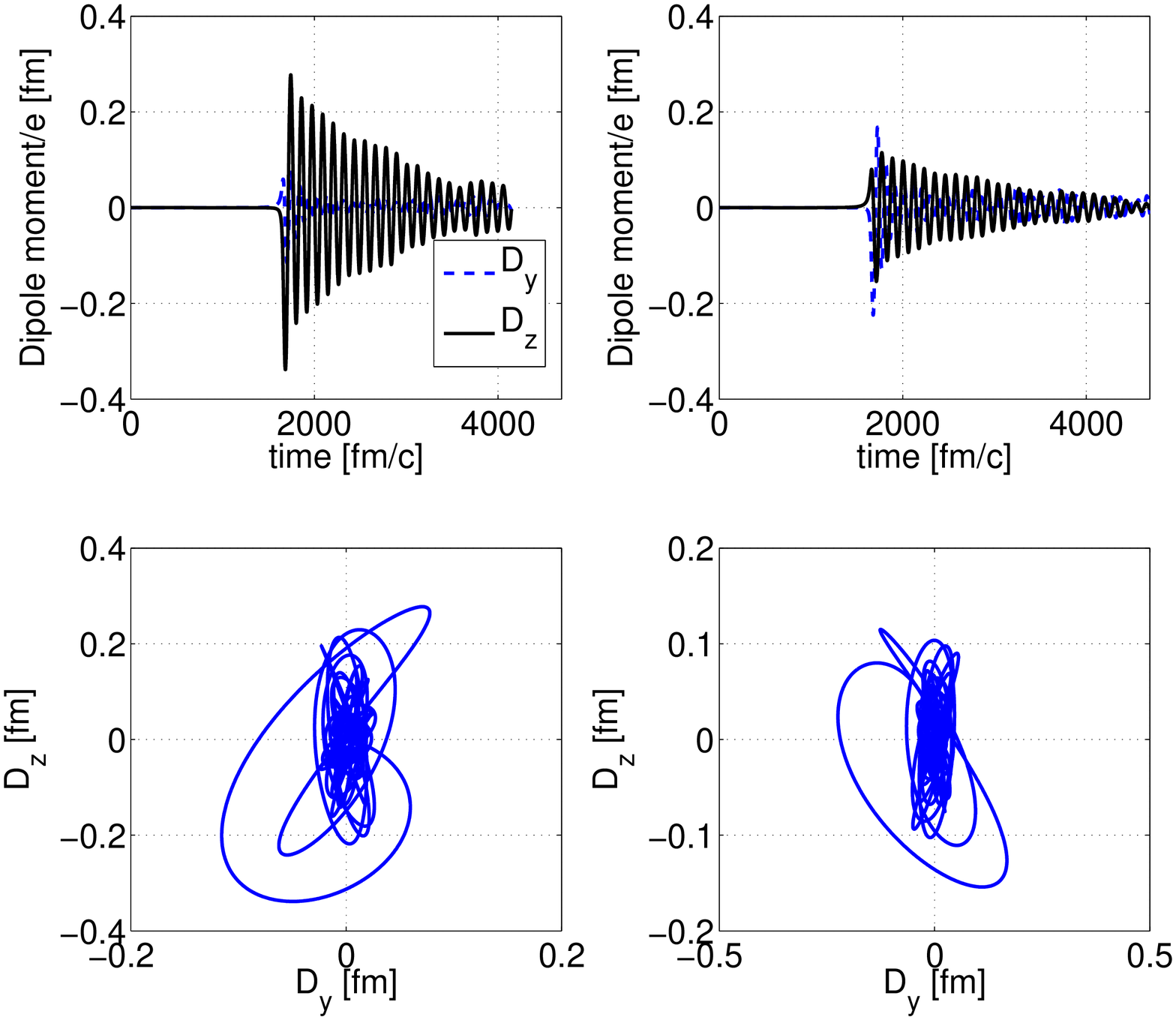}
\caption{ (color online) 
The same as in the Fig. \ref{fig7s}, but for the impact parameter $b=16.2$fm.
}\label{fig8s}
\end{figure}

\begin{figure} 
\includegraphics[width=9.0cm]{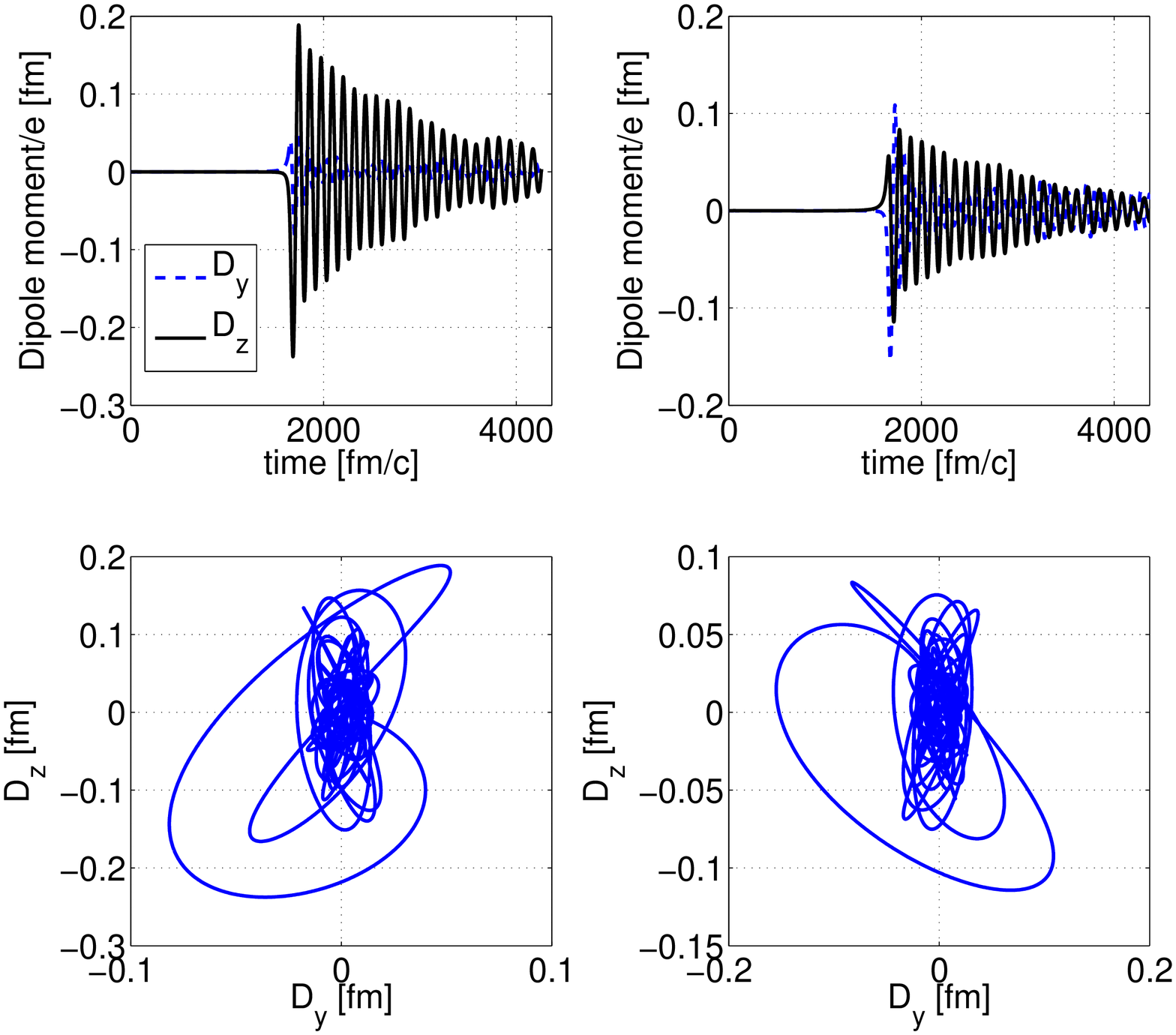}
\caption{ (color online) 
The same as in the Fig. \ref{fig7s}, but for the impact parameter $b=20.2$fm.
}\label{fig9s}
\end{figure}

The framework of TDSLDA allows to calculate various one body observables. In this case
the most important is the nuclear dipole moment. 
Only two components of the dipole moment, lying in the reaction plane, can oscillate as a result of collision.
In the Figs. \ref{fig7s}, \ref{fig8s}, \ref{fig9s}
these two components of the dipole moment have been plotted.


\begin{figure} 
\includegraphics[scale=0.4]{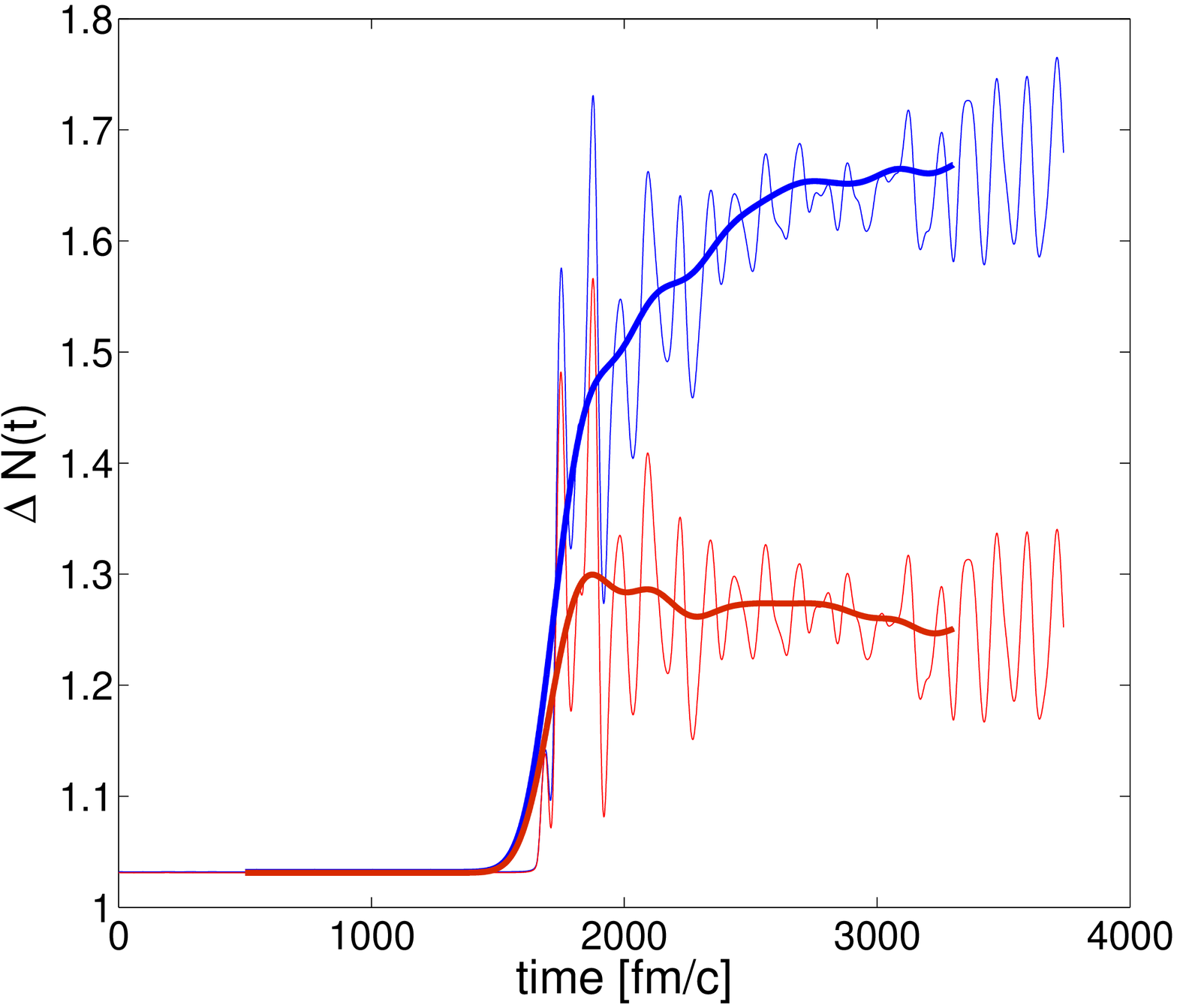}
\caption{ (color online) 
The number of neutrons present within the shell with inner radius $10$fm and outer radius $15$fm (red line).
 The number of neutrons present within the shell with inner radius $10$fm and outer radius $20$fm (blue line).
Thin line corresponds to the actual number of neutrons, whereas the thick line denotes the average value.
The plot corresponds to the collision with the target nucleus symmetry axis perpendicular to the trajectory
of the incoming projectile. The impact parameter $b=12.2$ fm. 
}\label{fig10sy}
\end{figure}

\begin{figure} 
\includegraphics[scale=0.4]{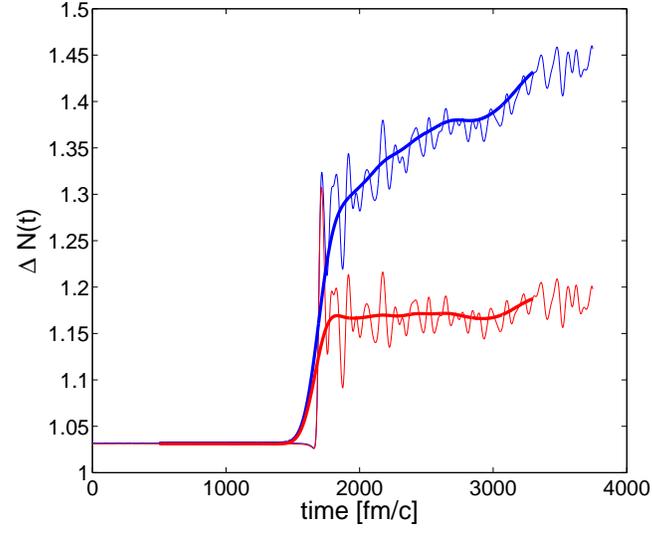}
\caption{ (color online) 
The same as in the Fig. \ref{fig10sy}, but for the nuclear orientation parallel with respect to the incoming
projectile.
}\label{fig10sz}
\end{figure}

\begin{figure} 
\includegraphics[scale=0.4]{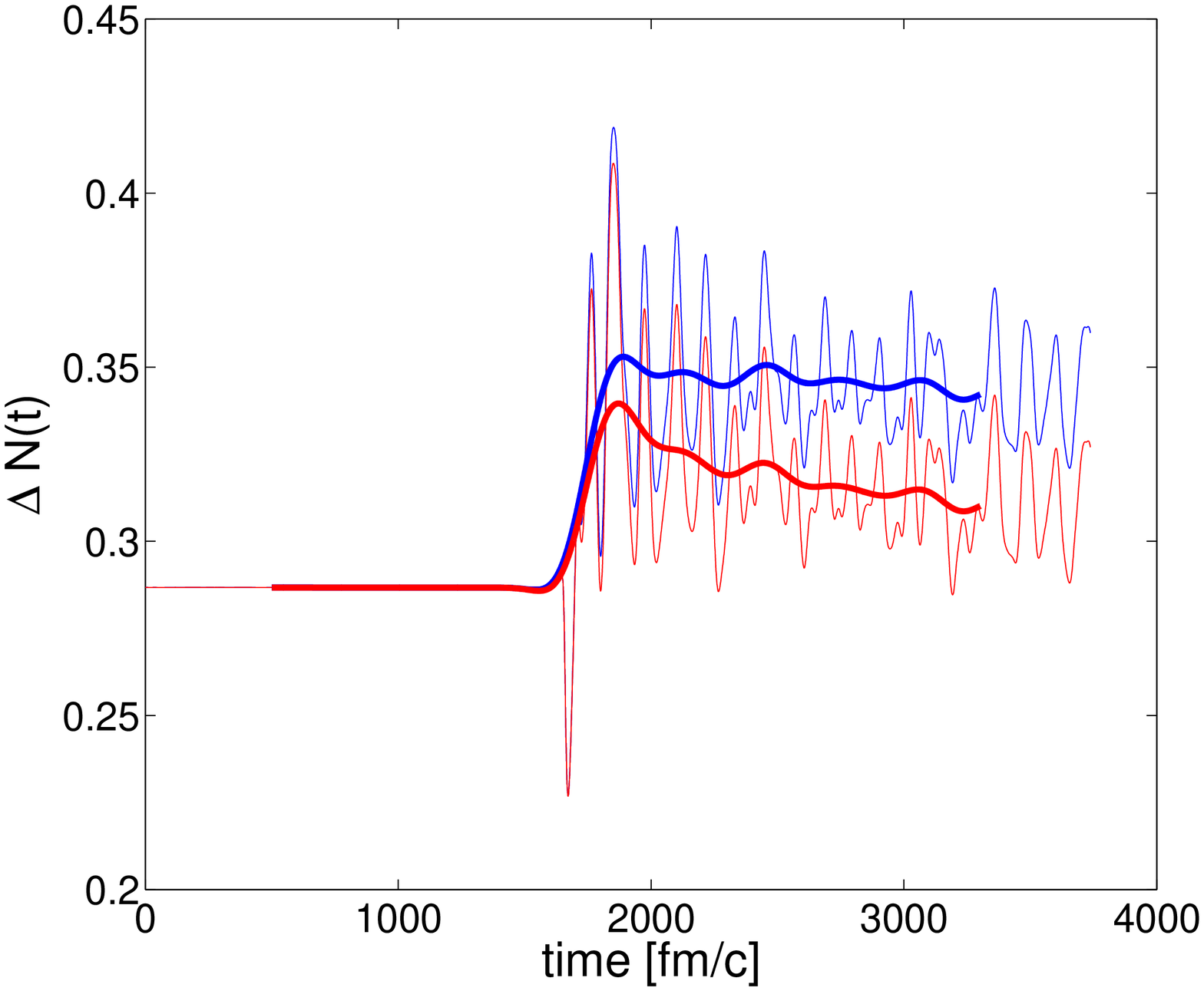}
\caption{ (color online) 
The same as in the Fig. \protect\ref{fig10sy}, but for protons.
}\label{fig11sy}
\end{figure}

\begin{figure} 
\includegraphics[scale=0.4]{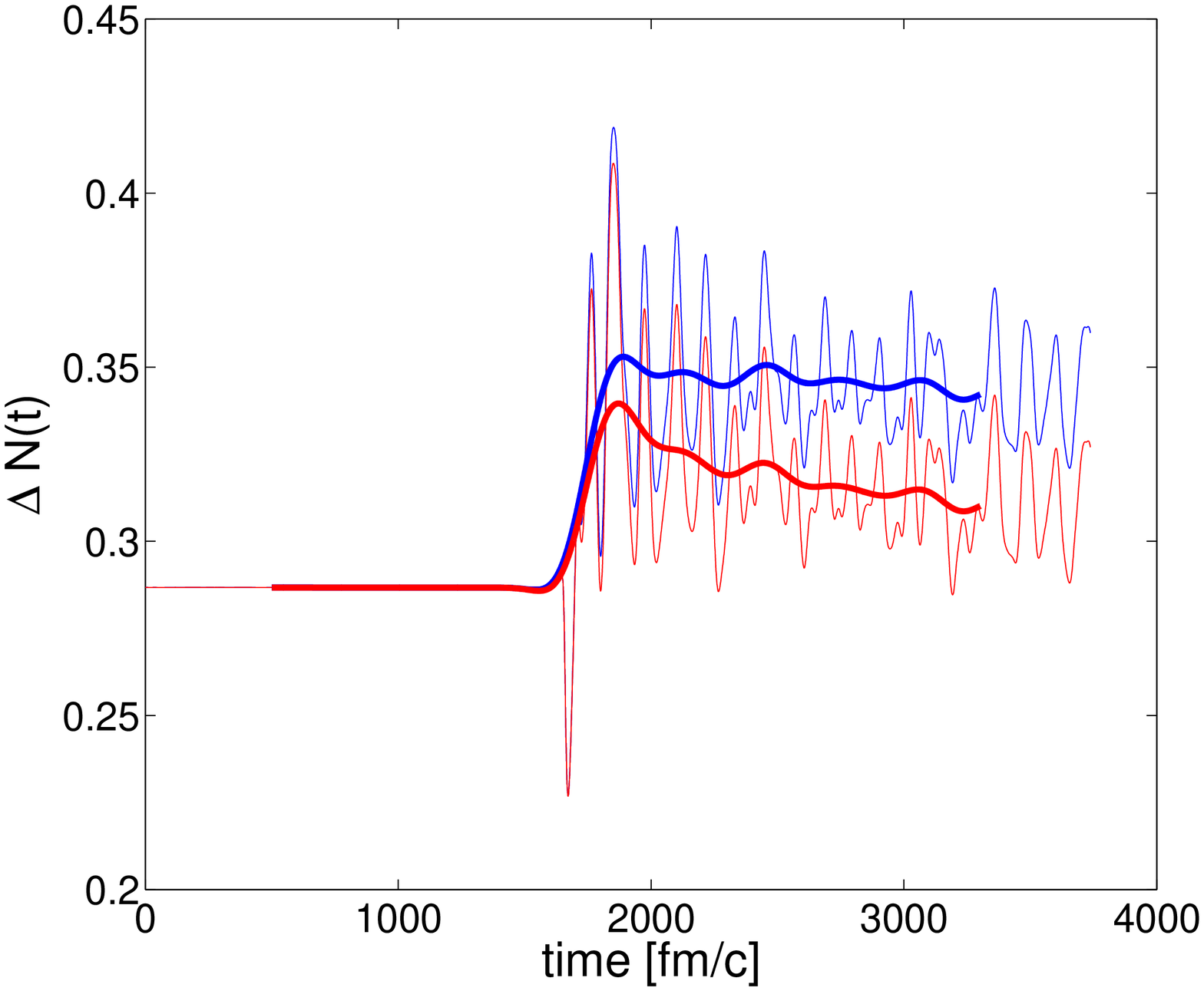}
\caption{ (color online) 
The same as in the Fig. \protect\ref{fig10sz}, but for protons.
}\label{fig11sz}
\end{figure}

During the time evolution the nucleus can emit particles.
In order to investigate this effect we have calculated the number of neutrons/protons
within shells of various radii. As one can see from the Figs. \ref{fig10sy}, \ref{fig10sz}, \ref{fig11sy}, \ref{fig11sz}
the number of protons in the shells outside the nucleus is negligible. Moreover this proton number
is approximately constant which indicate that we rather probe the tale of the proton distribution than
the emission process. On the contrary the situation is different for neutrons. The number of neutrons in the smaller
shell is much larger, although it is also approximately constant. However in the larger shell the number of neutrons 
is constantly increasing in time with a fairly constant average rate. It indicates that the neutron emission occurs as
a result of Coulomb excitation process. 

\end{document}